\documentclass[12pt]{iopart}
\usepackage[english]{babel}
\usepackage{
amssymb,amsfonts,latexsym}
\usepackage{graphicx}
\usepackage{color}
\usepackage{iopams}
\usepackage{subfigure}
\usepackage{afterpage}
\usepackage{blkarray}
\usepackage{cite}
\usepackage{tikz}
\usetikzlibrary{shapes.misc}
\usetikzlibrary{fadings}
\usetikzlibrary{arrows.meta}
\usetikzlibrary{decorations.pathmorphing}
\tikzset{cross/.style={cross out, draw=black, minimum size=2*(#1-\pgflinewidth), inner sep=0pt, outer sep=0pt},
cross/.default={1pt}}
\usetikzlibrary{calc}
\usepackage{bm}

\definecolor{blue(pigment)}{rgb}{0.2, 0.2, 0.6}
\usepackage{mathrsfs}
\usepackage{times}
\usepackage[colorlinks,
citecolor=blue,linkcolor=blue,urlcolor=blue]{hyperref}
\usepackage{scalerel}
\usepackage{tensor}
\usepackage{etoolbox}

\makeatletter
\def\@mkboth#1#2{}
\newlength\appendixwidth
\preto\appendix{\addtocontents{toc}{\protect\patchl@section}}
\newcommand{\patchl@section}{%
  \settowidth{\appendixwidth}{\textbf{Appendix }}%
  \addtolength{\appendixwidth}{1.5em}%
  \patchcmd{\l@section}{1.5em}{\appendixwidth}{}{\ddt}%
}
\makeatother

\newcommand{\eps}{\varepsilon}    
\def\eqref#1{(\ref{#1})}

\def\ket#1{|#1\rangle}

\def\bra#1{\langle#1|}

\usepackage{cancel}

\newcommand{\be}{\begin{equation}}
\newcommand{\ee}{\end{equation}}
\newcommand{\bea}{\begin{eqnarray}}
\newcommand{\eea}{\end{eqnarray}}

\newcommand\reallywidehat[1]{\arraycolsep=0pt\relax%
\begin{array}{c}
\stretchto{
  \scaleto{
    \scalerel*[\widthof{\ensuremath{#1}}]{\kern-.5pt\bigwedge\kern-.5pt}
    {\rule[-\textheight/2]{1ex}{\textheight}} 
  }{\textheight} %
}{0.5ex}\\           
#1\\                 
\rule{-1ex}{0ex}
\end{array}
}

\begin{document}

\title[What is an integrable quench?]{What is an integrable quench?}
\author{Lorenzo Piroli$^{1}$, Bal\'{a}zs Pozsgay$^{2,3}$, Eric Vernier$^{1}$}
\address{$ˆ1$ SISSA and INFN, via Bonomea 265, 34136 Trieste, Italy}
\address{$^2$ Department of Theoretical Physics, Budapest University
	of Technology and Economics, 1111 Budapest, Budafoki \'{u}t 8, Hungary}
\address{$ˆ3$ BME Statistical Field Theory Research Group, Institute of Physics,
	Budapest University of Technology and Economics, H-1111 Budapest, Hungary}

\ead{lpiroli@sissa.it, pozsgay.balazs@gmail.com, evernier@sissa.it}
\date{\today}

\begin{abstract}
Inspired by classical results in integrable boundary quantum field
theory, we propose a definition of integrable initial states for
quantum quenches in lattice models. They are defined as the states
which are annihilated by all local conserved charges that are odd under space reflection. We show that this class includes the states which can be related to integrable boundary conditions in an appropriate rotated channel, in loose analogy with the picture in quantum field theory. Furthermore, we provide an efficient method to test integrability of given initial states. We revisit the recent literature
of global quenches in several models and show that, in all of the cases where closed-form analytical results could be
obtained, the initial state is integrable according to our
definition. In the prototypical example of the XXZ spin-$s$ chains we show that integrable states include
two-site product states but also larger families of matrix product
states with arbitrary bond dimension. We argue that our results could be practically useful for the study of quantum quenches in generic integrable models.
\end{abstract}


\maketitle


\section{Introduction}\label{sec:intro}

The notion of solvability associated to integrable models usually refers to the possibility of diagonalizing analytically the corresponding Hamiltonian \cite{baxter-82}. Although this is a remarkable property, comparison with experiments often requires to go beyond the simple computation of the system's eigenspectrum and to provide predictions for more general physical observables. Indeed, a tremendous effort has been made in the past decade to compute several non-trivial ground-state and thermal properties in different models, significantly improving our understanding of the underlying mathematical structures \cite{korepin,JimboBOOK,takahashi,efgk-05,JiMi81,Slav89,JMMN92,MaSa96,KiMT00,
	GoKS04,CaMa05,BJMS07,TrGK09,CaCa06,KoMT10}.

More recently, integrable models offered new challenges to the community, as an increasing attention was devoted to the study of quantum quenches \cite{cc-06}. In this framework, one is interested in the unitary time evolution from an assigned initial state. It is evident that this problem is generally more complicated than those arising at thermal equilibrium, mainly because the initial state itself might be chosen arbitrarily, with an exponentially large number of degrees of freedom. This freedom hardly combines with the intrinsic rigidity of integrability, and one might legitimately wonder whether quench problems are within the reach of integrability-based techniques at all \cite{Delf14}. On the other hand, nearly ideal, out-of-equilibrium integrable models can now be realized experimentally in cold-atomic laboratories \cite{bdz-08,ccgo-11,pssv-11,LaGS16}, elevating the relevance of these questions beyond a purely academic curiosity (see also \cite{CaEM16} for a volume of reviews on integrable models out of equilibrium).

Among the most elementary quantities appearing in the study of quantum quenches are the overlaps between the initial state and the eigenstates of the model. These are needed as intermediate building blocks for several calculations both at finite and infinite times. Ever since the early stages of the literature on quantum quenches, several works tackled the problem of their computation, with the first studies dating back to almost ten years ago \cite{FaCC09,MoPC10}. However, until recently the problem appeared too hard to be attacked, strongly limiting our possibility of analytical calculations.

A breakthrough has come with the works \cite{KoPo12, Pozs14,BNWC14, DWBC14}, where non-trivial overlap formulas were obtained in the XXZ Heisenberg chain and in the Lieb-Liniger model. These formulas involve Gaudin-like determinants, and are reminiscent of the one for the norm of Bethe states \cite{Kore82}. After the seminal works \cite{KoPo12, Pozs14,BNWC14, DWBC14}, a large number of studies on the subject have appeared \cite{SoTM14,PiCa14,BDWC14-2,Broc14,LeKZ15,Bucc16,HoST16,
XX-quench-brockmann,BLKZ16,FoZa16,LeKM16,HoTa17,BrSt17}, and overlaps were computed for several initial states and models. In turn, these achievements finally allowed for the derivation of important analytical results by means of integrability-based methods such as the Quench Action approach \cite{CaEs13,Caux16}, which has already been applied to a large number of problems \cite{DWBC14,Bucc16,WDBF14,MPTW15,BeSE14,DeCa14,DeMV15,DePC15,BePC16,PiCE16,AlCa16,MBPC17,PiCa17,AlCa17-QR,BeTC17}. Quite remarkably, for all the known cases the overlap formulas display the same form, seemingly indicating the presence of a structure common to different models. However, so far it was not clear for which states the existence of such formulas could be expected. 

Paralleling these studies, analytical results for the post-quench time evolution were also obtained in cases where the overlaps are not known. In the works \cite{Pozs13,PiPV17} an exact computation of the Loschmidt echo was performed in the XXZ chain, starting from arbitrary families of two-site product states. These results could be derived by means of a construction relating the initial state to an integrable boundary condition in an appropriate rotated channel, where the space and time directions are exchanged. In fact, the connection between quantum quenches in one spatial dimension and boundary integrable quantum field theory (QFT) in the rotated channel has been known and exploited for a long time both in the conformal \cite{cc-06,CaCa05,CaCa16,Card16-2,Card17-RG} and massive cases \cite{Ghos94,GhZa94,FiMu10,SoFM12,BeSE14,PaSo14}.

This connection was in particular beautifully illustrated in the classical work of Ghoshal and Zamolodchikov \cite{GhZa94}. Here, integrability of the boundary field theory was defined by the existence of an infinite number of conserved operators (or charges) which persist after the addition of a boundary term to the bulk Hamiltonian. Remarkably, the conditions on this boundary term to preserve integrability were explicitly translated into a constraint for the boundary state in the corresponding rotated picture: the latter has to be annihilated by an appropriately chosen (infinite) subset of the bulk conserved charges.  

Inspired by these classical works, we propose a definition of integrable states for quantum quenches in lattice integrable systems. We identify integrable states as those which are annihilated by all local conserved charges of the Hamiltonian
that are odd under space reflection. This definition naturally extends to those models defined on the continuum that are scaling limit of lattice ones. We introduce an efficient method to test integrability of given initial states. By means of the latter, we show that in all of the cases recently considered where closed-form analytical results could be obtained, the initial state is integrable according to our definition.

Going further, we prove that integrable states include the ones which can be related to integrable boundaries in the rotated channel, where the space and time directions are exchanged \cite{PiPV17}. This result, which completes the analogy with the picture in QFT, is highly non-trivial, as the lattice model does not display Lorentz invariance. We devote a detailed analysis to the prototypical case of XXZ spin-$s$ chains and show that integrable states include two-site product states but also larger families of matrix product states (MPSs) with arbitrary bond-dimension. A particular case of the latter are the states studied in \cite{BLKZ16,LeKZ15,FoZa16}. 

It is important to stress that our findings should not be interpreted
as a no-go theorem for obtaining exact results for non-integrable
initial states. For example, an exact overlap formula was computed in
\cite{MoPC10} for the special case of the domain-wall state, which
strictly speaking is not integrable. Further analytical results were
obtained recently for quenches from the same state, for example
regarding the computation of the return probability \cite{Step17}, and
in the context of spin transport \cite{CoDV17}. However, these
problems are inherently inhomogeneous in space, and in the present
work we focus on the homogeneous global quenches.

Regarding global problems, relevant examples concern the computation of the post-quench rapidity distribution functions of the quasi-particles in XXZ spin-$s$ chains. Indeed, the latter can be computed exactly in many (non-integrable) cases \cite{IQNB16}. This has been a recent important achievement, based on a mature understanding of the physical relevance of quasi-local charges in integrable models \cite{IMPZ16,Pros11,PPSA14,PrIl13,IlMP15,Fago14,PiVe16,DeCD16,Fago17,VeCo17} and mainly motivated by studies regarding the validity of the so-called Generalized Gibbs Ensemble \cite{RDYO07,ViRi16,EsFa16,FaEs13,Muss13,
	Pozs13-GGE,FCEC14,Pozs14-failure_GGE,GoAn14,IDWC15,EsMP15,EsMP17,IlQC17,PiVC16,PoVW17}. However, even in this problem further simplifications occur for integrable states, allowing one to reach closed-form analytical formulas, via the so-called $Y$-system \cite{IQNB16,PiVC16}. In any case, our work provides a unified point of view for the
many exact results that appeared in the past years. Furthermore, it also provides a useful starting point for the study of models where quench problems have not been investigated: if exact results are to be derived, one should first look at the integrable initial states, for which they are expected.

The organization of this article is as follows. In Sec.~\ref{sec:general_discussion} we review the classical work of Ghoshal and Zamolodchikov and present the general setting. Integrable states are defined in Sec.~\ref{sec:def_integrable_states}, where we also discuss some of their properties. In Sec.~\ref{sec:integrable_boundaries} we show that states related to integrable boundaries in the rotated channel are integrable according to our definition, while in Sec.~\ref{sec:MPS} we present a general construction of integrable matrix product states with arbitrary bond dimension. Sec.~\ref{sec:integrable_qunches} is devoted to a critical review of several models considered in the recent literature. Finally, our conclusions are reported in Sec.~\ref{sec:conclusions}.

\section{General setting}\label{sec:general_discussion}

\subsection{Boundary states in integrable quantum field theory}\label{sec:boundary_QFT}

A source of inspiration for the present paper is the classical work of Ghoshal and Zamolodchikov \cite{GhZa94}, where integrable QFTs in the presence of boundaries are studied. It is natural to start our discussion by briefly reviewing the aspects of that work that are directly relevant for us. Although our focus will be on lattice integrable models, this will allow us to introduce some constructions by analogy with the picture in boundary QFT.

We start with an Euclidean field theory defined on a semi-infinite
plane $x\in(-\infty, 0)$, $y\in(-\infty, +\infty)$. In the so-called
Lagrangian approach it can be defined by introducing the corresponding
action, which generally reads as 
\be
\mathcal{A}_B=\int_{-\infty}^{+\infty}dy\int_{-\infty}^{0}dx\  a(\varphi,\partial_{\mu}\varphi)+\int_{-\infty}^{+\infty}dy\ b\left(\varphi_{\rm B},\frac{d}{dy}\varphi_{\rm B}\right)\,.
\label{eq:action_QFT}
\ee
Here $\varphi(x,y)$ represent a set of local \emph{bulk} fields, while $\varphi_{\rm B}(y)= \varphi_{\rm B}(x,y)|_{x=0}$ are the \emph{boundary} fields. Analogously, $a(\varphi,\partial_{\mu}\varphi)$ and $b\left(\varphi_{\rm B},d\varphi_{\rm B}/dy\right)$ are respectively the bulk and boundary action densities. 

\begin{figure}
\begin{tikzpicture}[scale=0.975]
\fill [blue!80!black,path fading=west] (0,0) rectangle (5,5); 
\draw[line width =4] (5,-0.5) -- (5,5.5);
\draw[dashed,line width =2, draw=black] (-0.25,2.5) -- (5,2.5);
\draw[-latex, line width =2, draw=blue!50!white] (5.8,-0.5) -- (5.8,5.5);
\node at (5.9, 5.7) {$t=\infty$};
\node at (6.15, -.7) {$t=-\infty$};
\node at (-0.1, 5.5) {$(a)$};
\draw[-latex, line width =2, draw=blue!50!white] (14.7,0) -- (14.7,5.5);
\fill [blue!80!black,path fading=north] (8.7,0) rectangle (13.7,5);
\draw[line width =4, draw=black] (8.2,0) -- (14.2,0);
\draw[line width =4, draw=blue!50!white] (14.45,0) -- (14.95,0);
\node at (14.8, 5.7) {$t=\infty$};
\node at (14.7, -0.5) {$t=0$};
\node at (8.6,5.5) {$(b)$};
\node at (11.2,-.5) {$|B\rangle$};
\end{tikzpicture}
\caption{Pictorial representation of a 2D Euclidean field theory in the presence of boundaries. There are two natural ways to introduce the Hamiltonian picture. In the first one, displayed in the sub-figure ~$(a)$, the Euclidean time direction is chosen to be parallel to the boundary. The physical Hilbert space is associated with a semi-infinite spatial line at fixed time (dashed black line). In the second way, displayed in the sub-figure $(b)$, the time direction is chosen to be orthogonal to the boundary. The latter plays the role of an initial condition, and is identified with a boundary initial state $|B\rangle$.}
\label{fig:2d_euclidean_QFT}
\end{figure}
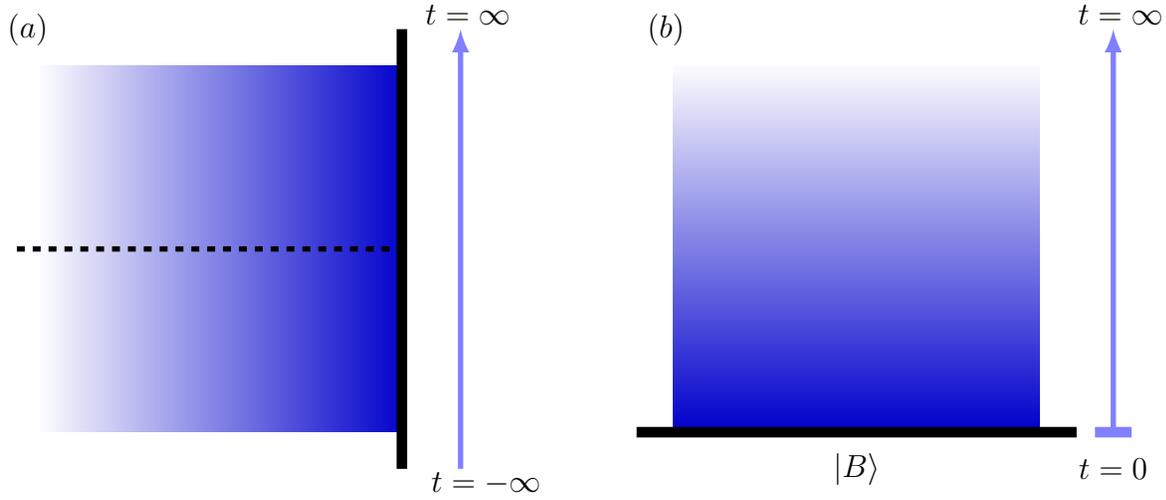

We recall that from the action \eqref{eq:action_QFT} one can introduce
a Hamiltonian picture in Euclidean time, which is closely related to
a 1+1 Lorentzian QFT through the Wick rotation.
This procedure of analytic continuation between real and imaginary
times is routinely performed
in order to apply QFT results to equilibrium statistical mechanics
\cite{Delf04}. As pointed out in \cite{GhZa94}, there are two natural
ways to introduce the Hamiltonian picture. This is pictorially
illustrated in Fig.~\ref{fig:2d_euclidean_QFT}. First, one can
identify the coordinate $y$ in \eqref{eq:action_QFT} to be the
Euclidean time direction. At each time, one can associate a physical
Hilbert space to the semi-infinite line $x\in (-\infty, 0)$. The
Hamiltonian is written as 
\be
H_{\rm B}=\int_{-\infty}^{0}dx\ \hat{h}(x) + \theta_{\rm B}\,,
\label{eq:boundary_hamiltonian_QFT}
\ee
where $\hat{h}(x)$ is the bulk Hamiltonian density, while $\theta_{\rm B}$ is an additional boundary term.

Alternatively, one can introduce an Hamiltonian picture in the \emph{rotated channel}, by identifying the time direction with the coordinate $x$. In this case, the Hilbert space at fixed time corresponds to the infinite line $y\in (-\infty,\infty)$ and the Hamiltonian coincides with the bulk one without boundary terms
\be
H=\int_{-\infty}^{+\infty}dy\ \hat{h}(y)\,.
\label{eq:bulk_hamiltonian_QFT}
\ee
The boundary at $x=0$ now plays the role of an initial condition in Euclidean time, and can be identified with an initial \emph{boundary} state $|B\rangle$, in analogy with the classical constructions in conformal field theory \cite{Card84}.

Consider now the bulk action density obtained from \eqref{eq:action_QFT} by removing the boundary. If the latter is integrable, there exist an infinite set of integrals of motions
\bea
P_{s}=\int_{-\infty}^{+\infty}dx\ \left[T_{s+1}+\Theta_{s-1}\right]\,,\quad 
\bar{P}_{s}=\int_{-\infty}^{+\infty}  dx \left[\bar{T}_{s+1}+\bar{\Theta}_{s-1}\right]\,\,,
\eea
where $T_{s+1}$, $\Theta_{s-1}$ ($\bar{T}_{s+1}$,
$\bar{\Theta}_{s-1}$) are local field of positive (negative) spin
$s+1$ and $s-1$, satisfying appropriate continuity equations
\cite{Zamo89}. The spin index $s$ takes value in an infinite subset
$\mathcal{S}$ of the positive integer numbers and it describes the
transformation properties under the Lorentz boosts.
Following
\cite{GhZa94}, a boundary field theory is defined to be integrable if
infinitely many conservation laws survive the addition of the boundary
term. More precisely, in this case the boundary Hamiltonian
\eqref{eq:boundary_hamiltonian_QFT} has an infinite number of integral
of motions of the form 
\be
H^{(s)}_{\rm B}=\int_{-\infty}^{0}dx\left[T_{s+1}+\bar{T}_{s+1}+\Theta_{s-1}+\bar{\Theta}_{s-1}\right]+\theta^{(s)}_{\rm B}\,.
\ee
The index $s$ now takes value in an infinite subset $\mathcal{S}_{\rm B}\subset \mathcal{S}$: the boundary integral of motions are obtained by selecting a subset of the bulk ones, and adding an appropriate boundary term to them. This is completely analogous to the situation encountered in lattice models with open boundary conditions \cite{GrMa96,Fago16}.

Remarkably, the existence of an infinite number of conservation laws for \eqref{eq:boundary_hamiltonian_QFT} can be related to a condition on the boundary state $|B\rangle$ in the rotated picture. If the Hamiltonian \eqref{eq:boundary_hamiltonian_QFT} is integrable then \cite{GhZa94}
\be
\left(P_{s}-\bar{P}_{s}\right)|B\rangle=0\,, \quad s\in \mathcal{S}_{\rm B}\,.
\label{eq:integrability_condition_QFT}
\ee
The relevance of \eqref{eq:integrability_condition_QFT} for our
purposes is that it represents a condition of integrability which can
be tested by relying uniquely on the knowledge of the initial boundary
state and the bulk conservation laws of the theory. This provides the
main source of inspiration for our work.


The application of the ideas of \cite{GhZa94} to lattice
models is not evident. A one dimensional quantum model is
always equivalent to a two-dimensional one of classical statistical physics, and
similar to the QFT setting it is always possible to define a ``rotated
channel''. However, typically there is no Euclidean invariance on the
2D lattice, and the Hamiltonians (or transfer matrices) of the rotated
channel differ from the original physical ones. Moreover, even if the identification of the initial states with the integrable boundary conditions of the rotated
channel can be carried out, this construction is non-trivial and in any case model dependent.

On the other hand, a definition analogous to \eqref{eq:integrability_condition_QFT} can be
straightforwardly introduced in a general way in lattice models, where
conservation laws are well known in terms of explicit operators on the
physical Hilbert space. In the next subsection we review some basic
facts on lattice integrable models which are necessary in order to
introduce and discuss our definition of integrable states. The latter will be
finally presented in Sec.~\ref{sec:def_integrable_states}.

\subsection{Lattice integrable models}\label{sec:general_setting}

We consider a generic one-dimensional model defined on the Hilbert space $\mathcal{H}=h_1\otimes \ldots \otimes h_L$. Here $h_j$ is a local $d$-dimensional Hilbert space associated with the site $j$, while $L$ is the spatial length of the system. The Hamiltonian is indicated as
\be
H_L=\sum_{j=1}^L \hat{h}_j\,,
\label{eq:generic_hamiltonian}
\ee
where $\hat{h}_j$ is a local operator. In the following, we will always assume periodic boundary conditions. In an integrable model there exists an infinite number of conserved operators commuting with the Hamiltonian \eqref{eq:generic_hamiltonian}. Furthermore, these can be written as a sum over the chain of local densities, and are therefore called local charges. The latter can be obtained by a standard construction within the so-called algebraic Bethe ansatz \cite{korepin,Fadd96}, which we now briefly sketch. 

One of the main objects of this formalism is the $R$-matrix $R_{1,2}(u)$. The latter is an operator acting on the tensor product of two local spaces $h_1\otimes h_2$ (possibly of different dimension), where $u$ is an arbitrary complex number, the spectral parameter. The $R$-matrix has to satisfy a set of non-linear relations known as Yang-Baxter equations
\be
R_{1,2}(u)R_{1,3}(u+v)R_{2,3}(v)=R_{2,3}(v)R_{1,3}(u+v)R_{1,2}(u)\,.
\label{eq:yang-baxter}
\ee
From the $R$-matrix, another fundamental object can be constructed, namely the transfer matrix 
\be
\tau(u)={\rm tr}_{0}\left\{T_0(u)\right\}\,,
\label{eq:transfer}
\ee
where the trace is over an auxiliary space $h_0$ and where we introduced the monodromy matrix
\be
T_0(u)=R_{0,L}(u)R_{0,L-1}(u)\ldots R_{0,1}(u)\,.
\label{eq:bulk_monodromy}
\ee
By means of \eqref{eq:yang-baxter}, one can show that transfer matrices with different spectral parameters commute
\be
[\tau(\lambda),\tau(\mu)]=0\,.
\label{eq:commutation}
\ee
Using \eqref{eq:commutation}, an infinite set of commuting operators is readily obtained as
\be
Q_{n+1}\propto  \frac{\partial^{n}}{\partial\lambda^{n}}\ln\tau(\lambda)\Big|_{\lambda=\lambda^*}\,\quad n\geq 1.
\label{eq:def_charges}
\ee
Importantly, $\lambda^*$ can be chosen in such a way that the
operators $Q_{n}$ are written as a sum over the chain of local
operators (or densities).
For example, within our conventions, the charges $Q_n$ are such that in the Heisenberg chains the corresponding
densities span $n$ sites.
Then, one can define an integrable Hamiltonian as
\be
H_L\propto Q_2\,.
\ee
By construction, $H_L$ is of the form \eqref{eq:generic_hamiltonian}, and has an infinite number of local charges $Q_n$. 

For the lattice models considered in this work, local conserved
charges can be divided into two subsets: the even ones, $Q_{2n}$, and
odd ones,  $Q_{2n+1}$ with $n\geq 1$. These sets display different
behavior under spatial reflection, namely\footnote{One of the simplest
  ways to prove the reflection properties is by using the so-called
  boost operator $B$, which connects the charges through the formal
  relation $Q_{k+1}=[B,Q_k]$ \cite{GM-higher-conserved-XXZ}. The boost operator is
  manifestly odd under space reflection, and this guarantees the
  alternating signs appearing in \eqref{eq:parity_charges}.
Note that the overall normalization of the transfer matrix affects the parity
properties of the charges: a rapidity dependent overall factor
introduces constant additive terms. However, in the models considered
in this work there is always a natural normalization in which 
the reflection properties \eqref{eq:parity_charges} hold.}
\be
\Pi Q_{k}\Pi=(-1)^{k}Q_{k}\,, \qquad k\geq 2\,,
\label{eq:parity_charges}
\ee
where $\Pi$ is the reflection operator
\be
\Pi|i_1, i_2, \ldots ,i_L\rangle =|i_L, i_{L-1}, \ldots ,i_{2}, i_1\rangle\,.
\label{eq:parity_operator}
\ee
Here we introduced the notation
\be
|i_1,i_2,\dots,i_L\rangle=|i_1\rangle_1 \otimes |i_2\rangle_2\ldots \otimes |i_L\rangle_L\,,
\label{eq:notation_vector_basis}
\ee
where $|k\rangle_j$ are the basis vectors of the local space $h_j$,
with $k=1,\ldots d$. 

We are interested in global quantum quenches where the Hamiltonian
driving the time evolution is integrable. For each model, we will
consider several families of initial states and in order to allow for
a general discussion, we will represent them as matrix product states (MPSs)
\cite{PVWC06}. They are a class of states which display a number of important properties: they have exponentially decaying correlations and finite entanglement between two
semi-infinite subsystems. Furthermore, it is known that MPSs can approximate ground states of gapped local Hamiltonians with arbitrary precision \cite{VeCi06}. This provides a physical motivation to consider them as initial states.

We recall that a generic (periodic) MPS can be defined as
\be
|\Psi_0\rangle=\sum_{i_1,\dots,i_L=1}^d{\rm tr}_{0}
\left[A_{1}^{(i_1)}  A_{2}^{(i_2)}  \dots A_L^{(i_L)}   \right]
|i_1,i_2,\dots,i_L\rangle\,.
\label{eq:generic_MPS}
\ee
Here $d$ is the dimension of the physical spaces $h_j$, while $A_j^{(i_j)}$ are $d_{j-1}\times d_{j}$ matrices, where $d_j$ are arbitrary positive integer numbers, called bond dimensions. The trace is over the Hilbert space $h_0$ with dimension $d_0$. In a finite chain, every vector of the Hilbert space $\mathcal{H}$ can be represented in the form \eqref{eq:generic_MPS} \cite{Vida03}. It is common practice, however, to refer to a state as MPS if the bond dimensions in the corresponding representation \eqref{eq:generic_MPS} do not grow with the system size $L$. 

Finally, it is useful to introduce the following definition: we say that a state $|\Psi_0\rangle$ is $p$-periodic if $p$ is the smallest positive integer such that
\be
U^{p}|\Psi_0\rangle=|\Psi_0\rangle\,,
\ee
where $U$ is the shift operator along the chain
\be
U|i_1, i_2, \ldots ,i_L\rangle =|i_L, i_1, \ldots ,i_{L-1}\rangle\,.
\label{eq:shift_operator}
\ee
In order to ensure a proper thermodynamic limit, we will restrict to initial states that are $p$-periodic,  with $p$ arbitrary but finite (and not increasing with $L$). The constraints imposed so far (namely, finite bond dimensions and $p$-periodicity) are extremely loose, and allow one to consider a very large family of initial states. Integrable states will be defined in the following as a small subset of the latter.

\section{Integrable states in lattice models}\label{sec:def_integrable_states}

\subsection{Defining integrable states}

We can now introduce our definition of integrable states, guided by
the analogy with the picture in QFT outlined
in Sec.~\ref{sec:boundary_QFT}. As we already stressed,
identification of initial states and boundary conditions in an
appropriate rotated channel requires preliminary work in lattice
models, and the construction can be model dependent.
However, it is possible
to introduce an immediate and general definition of integrable states in terms
of annihilation of a subset of the local conserved charges
$\{Q_n\}_{n=1}^{\infty}$, similarly to
Eq.~\eqref{eq:integrability_condition_QFT}.  

We propose the following definition: an initial state is integrable
if it is annihilated by all local charges of the model that are odd
under space reflection. 
In the lattice models
that we consider these coincide with the set $\{Q_{2n+1}\}$, with
$n\geq 1$, cf. Eq.~\eqref{eq:parity_charges}. Therefore we require
\be
Q_{2k+1}|\Psi_0\rangle =0\,, \qquad k\geq 1\,,
\label{eq:condition_integrability_lattice}
\ee
in any finite volume $L$ where the charge $Q_{2k+1}$ is well
defined. Typically $Q_n$ is a sum along the chain of local operators spanning $n$ sites, so that \eqref{eq:condition_integrability_lattice} is meaningful if $(2k+1)\le L$.

We stress that, even though the definition \eqref{eq:condition_integrability_lattice}
is inspired directly by \cite{GhZa94}, the analogy with QFT is a loose
one and its usefulness should therefore be
appreciated a posteriori. In particular, Eq.~\eqref{eq:condition_integrability_lattice} seems to hold for all the initial states for which closed-form analytical results could be obtained, at
least in the models considered in this work. Furthermore, we will see in Sec.~\ref{sec:integrable_boundaries} that the class of states satisfying \eqref{eq:condition_integrability_lattice} include all of those which can be
related to integrable boundaries in the rotated channel. These facts, together with additional considerations presented in the following, constitute strong evidence that \eqref{eq:condition_integrability_lattice} provides a meaningful and useful definition.


Based on the experience with the known cases we expect that in new
models and new quench situations analytic solutions can be found in
the integrable cases. Therefore, it is important to develop tools to
efficiently test the integrability of an initial state. In the
following we provide such methods, which
are connected directly to the definition
\eqref{eq:condition_integrability_lattice}, and are 
independent of the knowledge
of the overlaps or other composite objects such as the Loschmidt echo.

We consider a model defined by a transfer matrix $\tau(u)$, with the
so-called regularity condition
\be
\tau(0)=U\,,
\label{eq:regularity_conditions}
\ee
where $U$ is the translation operator
\eqref{eq:shift_operator}. Furthermore, we introduce the
constants of proportionality in \eqref{eq:def_charges} as
\be
Q_{n+1}=\alpha_n \frac{\partial^{n}}{\partial u^{n}}\ln\tau(u)\Big|_{u=0}\,,\qquad n\geq 1\,,
\label{eq:def_charges_norm}
\ee
where $\alpha_n$ are chosen such that the charges $Q_{n+1}$ are Hermitian. From this definition, one can write down the following formal representation
\be
\tau(u)=U\exp\left\{\sum_{n=1}^{\infty}\alpha_n\frac{u^n}{n!} Q_{n+1}\right\}\,.
\label{eq:formal_representation}
\ee
Integrability of a $p$-periodic MPS $|\Psi_0\rangle$ with finite bond dimension is equivalent to requiring
\be
\langle \Psi_0|Q_{2n+1}Q_{2n+1}|\Psi_0  \rangle=0\,,\qquad  n\geq 1\,,
\label{eq:vanishing_charges}
\ee
where we used that the charges are Hermitian operators.

In order to test \eqref{eq:vanishing_charges}, we introduce the quantities
\bea
G(u)&=&\frac{\langle \Psi_0|U^{p-2}\tau(u)\tau(-u)|\Psi_0 \rangle}{\langle\Psi_0|\Psi_0 \rangle}\,,\label{eq:g_functions}\\
\tilde{G}(u)&=&\frac{\langle \Psi_0|\Pi U^{p-2}\tau(u)\tau(-u) \Pi|\Psi_0 \rangle}{\langle\Psi_0|\Psi_0 \rangle}\,,
\label{eq:tilde_g_functions}
\eea
where $p$ is the periodicity of $|\Psi_0\rangle$. The motivation to
introduce a product of two transfer matrices is to cancel those parts
of the transfer matrix which are even and therefore irrelevant to the
integrability condition.

Our statement is that
\eqref{eq:vanishing_charges} holds if and only if in a neighborhood
of $u=0$
\be
G(u)=\tilde{G}(u)=1+O(u^L)\,.
\label{eq:integrability_test}
\ee
In the thermodynamic limit this leads to the condition
\be
G(u)=\tilde{G}(u)=1\,,\qquad |u|<K\,,
\label{eq:integrability_test_thermodynamic}
\ee 
for some $K>0$. Indeed, if the initial state is annihilated by all the odd charges in a given finite volume $L$, then the action of $\tau(u)\tau(-u)$ reduces to
$U^2+O(u^L)$ and we obtain \eqref{eq:integrability_test} and hence \eqref{eq:integrability_test_thermodynamic}. 
The proof of the other direction of the statement is also easy, but for the sake of clarity it is reported in \ref{sec:proof_integrability_condition}. 

Both functions $G(u)$ and $\tilde{G}(u)$ can be efficiently computed
on MPSs of finite bond dimension by standard techniques, as we will
also show explicitly in
Sec.~\ref{sec:transfer_matrix_calculations}. As a consequence,
Eqs.~\eqref{eq:integrability_test} and
\eqref{eq:integrability_test_thermodynamic} provide an efficient test
for integrability of given initial states. 

An alternative definition of integrability can be given
by requiring
\bea
\tau(u)|\Psi_0\rangle=\Pi \tau(u) \Pi|\Psi_0\rangle\,,
\label{eq:key_relation}
\eea
where $\Pi$ is the reflection operator \eqref{eq:parity_operator}. First, note that \eqref{eq:key_relation} directly implies two-site
 shift invariance:
 \bea
 U|\Psi_0\rangle=\tau(0)|\Psi_0\rangle=\Pi \tau(0) \Pi|\Psi_0\rangle
 =U^{-1}|\Psi_0\rangle\,,
 \label{eq:key_relation1}
 \eea
where we used $\tau(0)=U$. Next, the annihilation by the odd charges follows from
\eqref{eq:key_relation} simply by Taylor expanding $\tau(u)$ at $u=0$. Since \eqref{eq:key_relation} implies two-site shift invariance, it is a stronger condition than \eqref{eq:condition_integrability_lattice}. In fact, based on the analytic properties of the transfer matrix it can be
argued that \eqref{eq:key_relation} follows from
\eqref{eq:condition_integrability_lattice} and two-site
invariance. However, the question whether the latter property is
actually a 
consequence of the annihilation by all odd charges is an open problem.

\subsection{Transfer matrix evaluation of the integrability condition}\label{sec:transfer_matrix_calculations}

We consider a translational invariant MPS defined as
\begin{equation}
\ket{\Psi_0}=\sum_{i_1,\dots,i_L=1}^d{\rm tr}
\left[A^{(i_1)}  A^{(i_2)}  \dots A^{(i_L)}   \right]
\ket{i_1,i_2,\dots,i_L}\,,
\label{eq:translationally_invariant_MPS}
\end{equation}
and address the computation of the functions $G(u)$ and $\tilde G(u)$ introduced in the previous section. It is convenient to employ a graphical notation close to the one routinely used in the literature of tensor networks. This will help us to reduce the level of technicality of our discussions. First, we represent the $R$-matrix as
\be
\begin{tikzpicture}[baseline=(current  bounding  box.center)]
\node at (-1.7, 0.0) {$R_{0,1}(u) =$};
\draw (-0.3,0)  -- (1.3,0);
\draw (0.5,-0.8)  -- (0.5,0.8);
\node at (-0.6, 0.0) {$0$};
\node at (0.5, -1.2) {$1$};
\label{eq:representation_r_matrix}
\end{tikzpicture}
\ee
so that for the transfer matrix \eqref{eq:transfer} one has the graphical representation

\begin{equation}
\tau(u) =  ~~~
\begin{tikzpicture}[baseline=(current  bounding  box.center),xscale=1.5]
  \draw[black!50, dashed] (0,-2.5) arc (0:180:3 and 0.35); 
 \draw (0,-2.5) arc (0:-180:3 and 0.35);
 \draw[-latex, line width=1pt] (-3,-2.5)+(200:3 and 0.35) arc (200:205:3 and 0.35); 

\foreach \angle in {-150,-130,...,-20}
{ 
\draw ($(-3,-3)+(\angle:3 and 0.35)$)  -- ($(-3,-2)+(\angle:3 and 0.35)$) ;
}
\node at ($(-3.,-3.25)+(-150:3 and 0.35)$) {\small $1$};
\node at ($(-3.,-3.25)+(-130:3 and 0.35)$) {\small $2$};
\node at ($(-3.,-3.25)+(-90:3 and 0.35)$) {\ldots};
\node at ($(-3.,-3.25)+(-50:3 and 0.35)$) {\small $L-1$};
\node at ($(-3.,-3.25)+(-20:3 and 0.35)$) {\small $L$};

\label{eq:representation_transfer}
\end{tikzpicture}
~~~.
\end{equation}
The horizontal line denotes the auxiliary space $h_0$. Trace over $h_0$ is denoted with dashes, while vertical lines are are in 1-to-1 correspondence with the sites along the chain.

\begin{figure}
\centering
\begin{tikzpicture}[baseline=(current  bounding  box.center),scale=0.9]

\foreach \x in {-2.4,-1.6, -0.8,0} { 
	\draw (-1.2,\x)  -- (7.2,\x);
}

\foreach \x in {0,2,4,6} { 
		\draw (\x,-2.4)  -- (\x,0);
}

\foreach \x in {0,2,4,6} { 
	\filldraw[blue!40!white, draw=black] (\x-0.25,-2.4-0.25) rectangle (\x+0.25,-2.4+0.25);
}

\foreach \x in {0,2,4,6} { 
	\filldraw[blue!40!white, draw=black] (\x-0.25,-0.25) rectangle (\x+0.25,+0.25);
}
		
\foreach \x in {-2.4,-1.6, -0.8,0} { 
	\node[scale=1.2] at (-1.55,\x) {$\ldots$};	
}
	
\foreach \x in {-2.4,-1.6, -0.8,0} { 
	\node[scale=1.2] at (7.55,\x) {$\ldots$};	
}

\draw[rounded corners=10pt,line width=2] (3.3,0.0) -- (3.3,0.7) -- (4.7,0.7) -- (4.7,-3.1) -- (3.3,-3.1) -- (3.3, 0);

\node[scale=1.] at (5, 1) {$\tilde{\mathcal{T}}(u)$};	
\node[scale=1] at (8.4, -0.8) {$+u$};
\node[scale=1] at (8.4,-1.6) {$-u$};

\end{tikzpicture}
\caption{Graphical representation of Eq.~\eqref{eq:computation_G}. Filled boxes represent the matrices appearing in the MPS \eqref{eq:translationally_invariant_MPS}, whereas the internal lines stand for the insertion of two transfer matrices $\tau(u)$ and $\tau(-u)$, cf. \eqref{eq:representation_transfer}. The partition function of the resulting 2D lattice model can be se seen as generated by subsequent application of the matrix $\tilde{\mathcal{T}}(u)$ in the rotated channel.}
\label{fig:mpsqtm}
\end{figure}
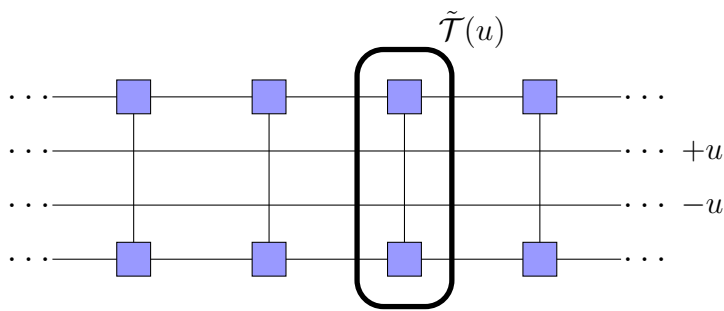

Using this notation the functions $G(u)$ and $\tilde{G}(u)$ in Eq.~\eqref{eq:g_functions} and \eqref{eq:tilde_g_functions} can be represented as partition functions of appropriate 2D statistical physical
models. This is pictorially depicted in Fig.~\ref{fig:mpsqtm}, where filled boxes
represent the matrices appearing in the MPS, whereas the internal lines stand
for the insertion of the transfer matrices $\tau(u)$ and $\tau(-u)$.

The partition function of the model displayed in Fig.~\ref{fig:mpsqtm} can be evaluated by building an alternative
transfer matrix (generally called the Quantum Transfer Matrix) which
acts in the horizontal direction. The calculations are standard, and analogous to those reported, for example, in Ref.~\cite{PiPV17}. In particular, it is easy to obtain 
\be
G(u)=\frac{1}{\mathcal{N}}{\rm tr}_{\mathcal{A}_1\otimes h_2 \otimes h_3\otimes \mathcal{A}_2}\left[\tilde{\mathcal{T}}(u)^{L}\right]\,,
\label{eq:computation_G}
\ee
where $\mathcal{N}=\langle\Psi_0|\Psi_0\rangle$ while
\be
\tilde{\mathcal{T}}(u)=\sum_{i=1}^d\sum_{j=1}^d\bar{A}^{(i)}\otimes 
\tensor[_{1}]{\langle i|}{}R_{2,1}(u)R_{3,1}(-u)|j\rangle_1 \otimes A^{(j)}\,,
\label{eq:monodromy_MPS}
\ee
and where $\mathcal{A}_1$ and $\mathcal{A}_2$ are $d$-dimensional spaces
on which the matrices $A^{(j)}$ act. Here $\bar{A}^{(i)}$ indicates the complex conjugated of $A^{(i)}$. An analogous computation can be
carried out for $\tilde{G}(u)$ and also for MPSs that are $p$-periodic.

The integrability conditions \eqref{eq:integrability_test} translate
into a constraint for the eigenvalues $\{\Lambda_j(u)\}_{j}$ 
of $\tilde{\mathcal{T}}(u)$, namely
\be
\frac{1}{\mathcal{N}}\left[\sum_{j}\left(\Lambda_j(u)\right)^L\right]\equiv 1 +O(u^L)\,.
\label{eq:test_explicit_eigen}
\ee
Comparing with the special point $u=0$ we obtain the norm as
\begin{equation*}
\mathcal{N}=\sum_{j}\left(\Lambda_j(0)\right)^L
\end{equation*}
It can be seen that \eqref{eq:test_explicit_eigen} is satisfied when
for $|u|<K$ with some $K\in\mathbb{R}^+$
\begin{equation}
  \label{eq:eigcond}
  \Lambda_j(u)\equiv\Lambda_j(0) \quad\textrm{for those}\ j\
  \textrm{for which}\quad |\Lambda_j(0)|>0.
\end{equation}
Condition \eqref{eq:eigcond} implies strict annihilation in finite
volume. If we only require asymptotic annihilation in the
thermodynamic limit then it is enough for \eqref{eq:eigcond} to hold
for the eigenvalues with maximal magnitude. However, we suggest to use
exact annihilation for the definition of integrability, and we will
show that all the previous cases studied in the literature fit this definition. 
  
It is important that for a given MPS the matrix
\eqref{eq:monodromy_MPS} has finite dimension (the latter growing with
the bond dimension of the MPS).  Accordingly its eigenvalues can be investigated either
analytically or numerically. Therefore, it is immediate to test
integrability of given MPSs.
In \ref{sec:examples_QTM_construction} we report two examples where the
integrability condition is tested both for an integrable and a
non-integrable initial state. 

It is a relevant question to find all integrable MPSs for a
given model. One strategy would be to write down a set of
conditions for the matrices $A^{(j)}$ appearing in \eqref{eq:translationally_invariant_MPS} such that the eigenvalues of the
Quantum Transfer Matrix \eqref{eq:monodromy_MPS} have the necessary properties. 
However, this approach appears to be extremely complicated and not viable. In general, we have not succeeded in solving this problem. Nevertheless, in the following
section we present
a general method to construct integrable MPSs of
arbitrary bond dimension. We will show later that the integrable MPSs studied in the recent literature all fit into our framework. The question whether all integrable MPSs can be generated by our method is left for future work.

\subsection{The pair structure}\label{sec:pair_structure}

The integrability condition \eqref{eq:condition_integrability_lattice}
has immediate consequences for the overlaps between integrable states and the eigenstates of the Hamiltonian (which in the following
will be also called Bethe states).
For a generic model, it is
well known that the latter can be parametrized by sets of
quasi-momenta or rapidities $\{\lambda_j\}_{j=1}^N$, where $N$ is the
number of quasi-particles associated to the eigenstate. Any initial
state can be written as 
\be
|\psi_0\rangle=\sum_{\{\lambda_j\}_j} c_{\{\lambda_j\}_j}|{\{\lambda_j\}_j}\rangle\,,
\ee
where the sum is over all the sets of rapidities, and where
$c_{\{\lambda_j\}_j}$ are the corresponding overlaps with the initial
state. 

Given a local charge $Q_{r}$, its action on Bethe states is
\be
Q_r|{\{\lambda_j\}_{j=1}^N}\rangle=\left[\sum_{j=1}^Nq_r(\lambda_j)\right]|{\{\lambda_j\}_{j=1}^N}\rangle\,,
\ee
where $q_r(\lambda)$ is some known function. As a consequence, an integrable state can have a non-vanishing overlap $c_{\{\lambda_j\}_j}$ only with the Bethe states $|{\{\lambda_j\}_j}\rangle$ such that 
\be
\sum_{j=1}^Nq_{2n+1}(\lambda_j)=0\,,
\label{eq:condition_overlap}
\ee
for all $n$ such that $Q_{2n+1}$ exists in the chain of length
$L$. In an interacting model this is a very strong constraint for the rapidities,
because they have to satisfy a set of additional quantization conditions known as Bethe equations. Accordingly, only special
configurations are consistent with \eqref{eq:condition_overlap}.

It follows from the space-reflection properties that the functions
$q_{2n+1}(\lambda)$ are odd with respect to $\lambda$.
Depending on the specific model chosen, there can be a finite set
$\mathcal{S}_{\lambda}$ of rapidity values such that
$q_{2n+1}(\lambda)=0$ if $\lambda\in
\mathcal{S}_{\lambda}$. Accordingly, the constraint
\eqref{eq:condition_overlap} is obviously satisfied by states
corresponding to a set 
\be
\{\lambda_j\}_{j=1}^N=\{\mu_j\}_{j=1}^R\cup \{-\mu_j\}_{j=1}^R\cup C\,,\quad C\subset \mathcal{S}_{\lambda}\,, \label{eq:pair_structure}
\ee
where $R$ is some positive integer number. Eq.~ \eqref{eq:pair_structure} encodes the so-called \emph{pair structure}. More precisely, we say that an initial state has the pair structure if it has non-vanishing overlap only with Bethe states corresponding to sets of rapidities of the form \eqref{eq:pair_structure}.

The presence of the pair structure was already observed in the seminal works \cite{DWBC14,BNWC14,Broc14,WDBF14,MPTW15}
 in the Lieb-Liniger and XXZ models, and has been repeatedly encountered in the recent literature for different initial states and systems. In the context of boundary QFT, it leads to the specific
``squeezed'' form of the boundary states 
\cite{GhZa94}, see
also \cite{FiMu10,SoTM14,HoST16,HoTa17,Pozs11,ScEs12,RiIg11,Evan13,KoZa16,CuSc17}. Furthermore, the pair structure has immediate consequences for the entropy of the
steady state arising after a quench \cite{AlCa17-QR,PVCR17,BeTC17}, in particular for its relation with the so-called diagonal entropy \cite{PVCR17,ReKo06,Polk11,SaPR11,FaCa08,Gura13,CoKC14}.
Therefore, it is an important question whether the pair structure follows in general from
\eqref{eq:condition_overlap}. In passing, we note that in the context
of QFT it has been argued that for interaction quenches near
criticality the pair structure does not occur. Namely the initial
state (the ground state of a critical Hamiltonian) does not consist
uniquely of pairs of particles with opposite momenta \cite{Delf14},
see also the related work \cite{Schu15}.

In a generic interacting model the Bethe
equations are algebraically independent of
\eqref{eq:condition_overlap} and a fine tuning of the couplings and/or
an interplay with the volume parameter is needed to find exceptions to
the pair structure. 
Such fine tuned examples 
can be found in the XXZ 
model at the special points $\Delta=\cos(p\pi/q)$, including the
free fermionic point $\Delta=0$, which will be treated in
more detail in section \ref{sec:xx_model}. In the case of the isotropic Heisenberg chain a rigorous proof of the
pair structure can be given using 
a result of \cite{tarasov-varchenko-xxx-simple}. Since the argument is
very simple, we present it here for completeness.

For an arbitrary Bethe state we have  the relation
$\Pi\ket{\{\lambda\}_N}\sim \ket{\{-\lambda\}_N}$. 
Taking scalar product of the two sides of the equation
\eqref{eq:key_relation} with the Bethe state $\ket{\{\lambda\}_N}$ we
see that the overlaps can be non-zero only if
\begin{equation}
  \tau(u,\{\lambda\}_N)= \tau(u,\{-\lambda\}_N)\,,
\end{equation}
for all $u$, where $\tau(u,\{\lambda\}_N)$ is the eigenvalue of the transfer matrix corresponding to $\ket{\{\lambda\}_N}$. It was proven in
\cite{tarasov-varchenko-xxx-simple} 
that the spectrum of the transfer matrix $\tau(u)$ is simple: if two different eigenstates share the
same eigenvalue then they belong to the
same $SU(2)$ multiplet.
This implies that $\ket{\{\lambda\}_N}$,
and $\Pi\ket{\{\lambda\}_N}$ are related by an $SU(2)$ rotation. Since $\Pi$ does not change the $S_z$
eigenvalues, we conclude that $\Pi\ket{\{\lambda\}_N}\sim
\ket{\{\lambda\}_N}$, and the pair structure holds.

To conclude this section, we stress that while the pair structure might or
might not hold for a specific model (and has to be investigated 
separately), the conditions
\eqref{eq:condition_integrability_lattice} and
\eqref{eq:condition_overlap} are general unifying properties of
integrable initial states.

\section{Relation with integrable boundaries}\label{sec:integrable_boundaries}

In the previous sections we have introduced the definition of
integrable states in terms of the bulk conserved charges of the
Hamiltonian. This definition was inspired by the picture in QFT, where integrable
boundaries are directly related to integrable initial states. It is
then a
natural question to ask, whether such a relation holds also in lattice
models, and whether our definition
\eqref{eq:condition_integrability_lattice} is compatible with it.

In this section we prove one direction of this relation:
we present a method to relate integrable boundary conditions to
initial states, and show that states obtained in this way 
indeed satisfy the condition
\eqref{eq:condition_integrability_lattice}. This construction naturally produces local two-site
product states, which are presented in Sec \ref{sec:loc}. In Section
\ref{sec:MPS} we show how integrable MPSs can also be taken into
account in this framework.

\subsection{The general construction}

\label{sec:loc}

The construction to relate integrable initial states to integrable
boundaries relies on the path integral evaluation of the partition
function
\be
\mathcal{Z}(w)=\langle \Psi_0|e^{-w H}|\Psi_0\rangle\,.
\label{eq:euclidean_partition_function}
\ee
In QFT the exchange of time and space directions is straightforward due to
the Euclidean invariance of the path integral. However, this is less
immediate in lattice models, where space is discrete and time is continuous. The standard method to circumvent this
problem is to introduce a discretization in the time direction and
then to develop a lattice path integral for the resulting partition function. This is achieved by employing a Trotter decomposition
%
\be
e^{-wH}\sim \left(1-\frac{w}{N}H\right)^{N}\,.
\label{eq:suzuki_trotter}
\ee
In integrable models the Hamiltonian $H$ can be related to the
transfer matrices, and this makes it possible to introduce the lattice
path integrals.

For the sake of clarity here we focus on the XXZ spin-$1/2$
  model, defined by the Hamiltonian
\bea
H_{L}&=& \frac{1}{4}\sum_{j=1}^{L}\left[\sigma^{x}_j\sigma^{x}_{j+1}+  \sigma^{y}_j\sigma^{y}_{j+1}+\Delta\left( \sigma^{z}_j\sigma^{z}_{j+1}-1\right)\right] \,,
\label{eq:hamiltonian_xxz}
\eea
where $\sigma_{j}^{\alpha}$ are the Pauli matrices, and periodic boundary conditions are assumed, $\sigma_{L+1}^{\alpha}=\sigma_{1}^{\alpha}$. In this case, the $R$-matrix is
\be  
R(u) 
= \frac{1}{\sinh\left(u+\eta\right)}
\left(
\begin{array}{cccc}
	\sinh(u+\eta) & & & \\
	& \sinh u & \sinh \eta & \\
	&\sinh \eta  & \sinh u  & \\
	& & & \sinh(u+\eta)
\end{array}
\right)
\,,
\label{eq:Rmatrix_XXZ}
\ee
where $\eta={\rm arccosh}(\Delta)$. The normalization of
  the $R$-matrix is such that the corresponding transfer matrix both
  satisfies the regularity condition \eqref{eq:regularity_conditions}
  and has charges $Q_n$ with the correct even/odd behavior
  \eqref{eq:parity_charges}. Moreover, it satisfies the so-called crossing
relation 
\be
R^{t_0}_{0,1}(u)=\gamma(u) V^{-1}_0R_{0,1}(-u-\eta)V_0\,.
\label{eq:crossin_relation}
\ee
Here $R^{t_0}_{0,1}(u)$ denotes partial transposition in the space
$h_0$, $V_0$ is a gauge matrix acting on $h_0$ while $\gamma(u)$ is a
function. In particular, for the $R$-matrix \eqref{eq:Rmatrix_XXZ} one
has $\gamma(u)=\sinh(u)/\sinh(u+\eta)$ and $V_0=\sigma^y$. While here we
 focus on the XXZ spin-$1/2$ chain, the construction described in
this section can be carried out straightforwardly for all the
integrable models whose $R$-matrix satisfies
a crossing relation, and is thus very general. An example is
given by higher spin versions of the Hamiltonian
\eqref{eq:hamiltonian_xxz}, as we will see in the
following. On the other hand, generalization of this construction to models for which
there is no relation of the type \eqref{eq:crossin_relation} (such as the
$SU(3)$-invariant spin chain) is not evident and needs further research.

We follow the derivation of \cite{PiPV17}, to which we refer for all the necessary technical details, providing here only the main ideas. For simplicity, we start by considering an initial state of the form
\be
|\Psi_0\rangle=|\psi_0\rangle_{1,2}\otimes \ldots |\psi_0\rangle_{L-1,L}\,,
\label{eq:initial_two_site}
\ee
while MPSs with arbitrary bond dimension will be considered in the next section. It is important to recall that, similarly to \eqref{eq:boundary_hamiltonian_QFT}, one can consider the
Hamiltonian \eqref{eq:hamiltonian_xxz} with open boundary conditions preserving integrability. In this case the open Hamiltonian can still be obtained by
means of an algebraic construction similar to the one described in
Sec.~\ref{sec:general_setting}. In the boundary case the relevant
object is a two-row transfer matrix, which reads
\be
\tau_{\rm B}(u)={\rm tr}_{0}\left\{K^{+}_0(u)T_0(u)K_0^{-}(u)\hat{T}_0(u)\right\}\,,
\label{eq:open_transfer}
\ee
where  
\bea
T_0(u)&=&R_{L,0}(u-\xi_L)\ldots R_{1,0}(u-\xi_1)\,,\\
\hat{T}_0(u)&=&R_{1,0}(u+\xi_1-\eta)\ldots R_{L,0}(u+\xi_L-\eta)\,,
\label{eq:general_integrable_boundary}
\eea
and where $\xi_j$ are arbitrary inhomogeneities. The $2\times 2$ matrices $K_0^{\pm}(u)$ are boundary operators acting on the auxiliary space $h_0$, which in this case read $K^{\pm}(u)=K(u\pm \eta/2,\xi_{\pm},\kappa_{\pm},\tau_{\pm})$, with 
\bea
\fl K(u,\xi,\kappa,\tau)= \left(\begin{array}{cc}
	k_{11}(u)&k_{12}(u) \\
	k_{21}(u)&k_{22}(u)
\end{array}\right) =
\left(\begin{array}{cc}
	\sinh\left(\xi+u\right)&\kappa e^{\tau} \sinh\left(2u\right) \\
	\kappa e^{-\tau}\sinh\left(2u\right)&\sinh(\xi-u)
\end{array}\right)\label{eq:k_matrix_nondiag}\,. 
\eea
Here $\xi$, $\kappa$ and $\tau$ are arbitrary parameters. The $K$-matrix has to satisfy the so-called reflection (or boundary Yang-Baxter) equations
\bea
R_{1,2}(u-w)K_{1}(u)R_{1,2}(u+w)K_{2}(w)=\nonumber\\
K_{2}(w)R_{1,2}(u+w)K_{1}(u)R_{1,2}(u-w)\,,
\label{eq:reflection_equation}
\eea
which guarantee the commutativity of the two-row transfer matrices \eqref{eq:open_transfer}.

We have now all the elements to evaluate the partition function \eqref{eq:euclidean_partition_function} from the Trotter decomposition \eqref{eq:suzuki_trotter}. In the particular
case of the XXZ Heisenberg chain, one can write \cite{Klum92,Klum04} 
\be
\left(1-\frac{w}{N}H\right)^{N}\sim \left[\tau(w^{*}/2N)\tau(\eta-w^{*}/2N)\right]^{N}\,,
\label{eq:discrete_time_evolution}
\ee
where $\tau(\lambda)$ is the usual periodic transfer matrix and
$w^{*}=(w\sinh\eta)/2$. Eqs.~\eqref{eq:suzuki_trotter} and
\eqref{eq:discrete_time_evolution} can be interpreted as follows: the
continuous time evolution can be approximated by a discrete one where
the elementary time step is obtained from the application of a two-row
transfer matrix. In order to parallel the discussion of
Sec.~\ref{sec:boundary_QFT}, we consider Euclidean time.

%

\begin{figure}
	\begin{center}
		\begin{tikzpicture}[scale=1.1]

\foreach \y in {0,1.25,...,3}
{ 

  \draw[black!40, dashed] (8,\y) arc (0:180:5 and 0.5); 
 \draw[] (8,\y) arc (0:-180:5 and 0.5); 
}

\foreach \angle in {-150,-130,...,-20}
{ 
\draw[dashed] ($(3,4)+(\angle-10:5 and 0.5)$) -- ($(3,3.)+(\angle-10:5 and 0.5)$);
\draw[dashed] ($(3,4)+(\angle:5 and 0.5)$) -- ($(3,3.)+(\angle:5 and 0.5)$);

\draw[rounded corners=1pt] 
($(3,3)+(\angle-10:5 and 0.5)$) -- ($(3,-0.5)+(\angle-10:5 and 0.5)$)  --  ($(3,-0.75)+(\angle-5:5 and 0.5)$)  -- ($(3,-0.5)+(\angle:5 and 0.5)$)  -- ($(3,3)+(\angle:5 and 0.5)$);

\node at ($(3.,-1)+(\angle-5:5 and 0.5)$) {\small $\ket{\psi_0}$};
}

		\draw[-latex, line width =2, draw=blue!50!white] (9,-1) -- (9,4.);
		\node at (9.7, 2.0) {$t$};
		\draw[line width =4, draw=blue!50!white] (8.75,-1) -- (9.25,-1);
		\node at (9, 4.4) {$t=\infty$};
		\node at (9, -1.3) {$t=0$};
		\end{tikzpicture}
		\caption{Pictorial representation of the quench dynamics from an integrable state, after discretization of the Euclidean time direction. Parallel rows indicate the action of product of transfer matrices, which implement the unit-step along the Euclidean time direction. }
		\label{fig:2d_classical}
	\end{center}
\end{figure}
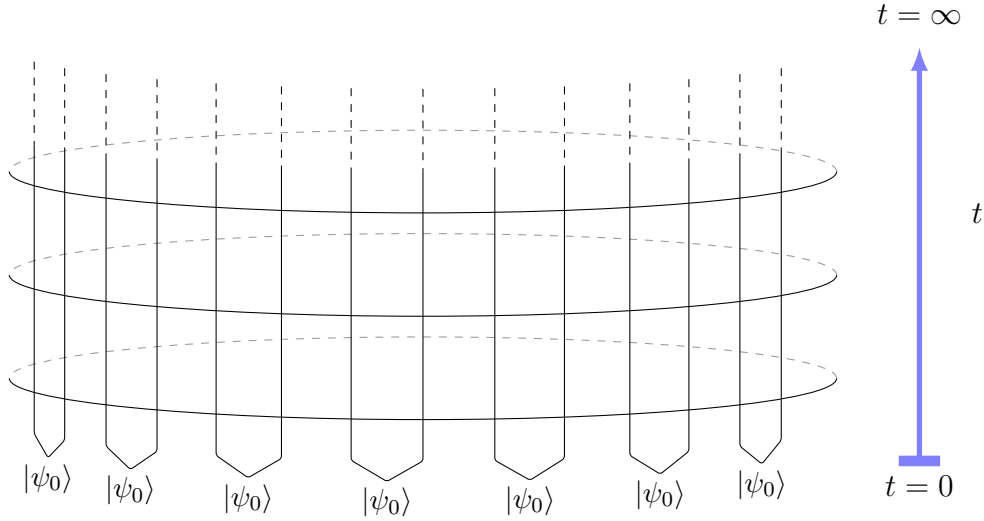

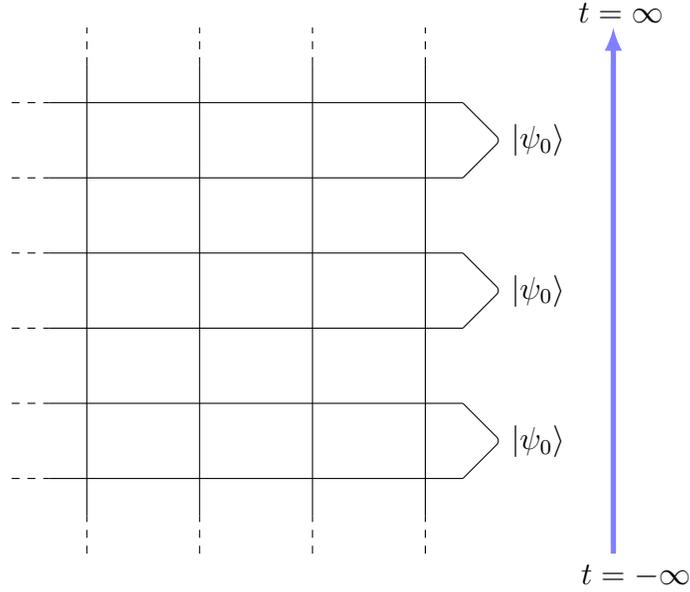
\begin{figure}
	\begin{center}
		\begin{tikzpicture}
		\foreach \x in {1,2.5,4}{
			\draw (\x, -2) -- (\x, 4);
		}
		\draw (5.5,-2)  -- (5.5,4);
		\foreach \x in {-1.5,-0.5,0.5,...,3.5}{
			\draw (0.5,\x) -- (6, \x);
			\draw[dashed] (0,\x) -- (0.5,\x);
		};
		\foreach \x in {-1.5,0.5,...,2.5}{
			\draw[rounded corners=2pt] (6,\x) -- (6.25,\x+0.25)  -- (6.5,\x+0.5) --  
			(6.25, \x+0.75) -- (6,\x+1);
			\node at (7,\x+0.5) {$|\psi_0\rangle $};
		}
		\foreach \x in {1,2.5,4}{
			\draw[dashed] (\x,-2.5) -- (\x,-2);
			\draw[dashed] (\x,4) -- (\x,4.5);
		};
		\draw[dashed] (5.5,-2.5) -- (5.5,-2);
		\draw[dashed] (5.5,4) -- (5.5,4.5);
		\draw[-latex, line width =2, draw=blue!50!white] (8,-2.5) -- (8,4.5);
		\node at (8.1, 4.7) {$t=\infty$};
		\node at (8.3, -2.8) {$t=-\infty$};
		
		\end{tikzpicture}
		\caption{Pictorial representation of the quench from an integrable state in the rotated channel. By choosing the Euclidean time direction along the boundary, in analogy with Fig.~\ref{fig:2d_euclidean_QFT}, the initial state is interpreted as a boundary in space. The Euclidean time evolution can be thought of as generated by an open transfer matrix in which the reflection matrix is specified by the initial state. For particular choices of the latter the open transfer matrix becomes integrable. }
		\label{fig:2d_classical_rotated_channel}
	\end{center}
\end{figure}

From the Trotter-Suzuki decomposition \eqref{eq:suzuki_trotter} and \eqref{eq:discrete_time_evolution} it is evident that the computation of \eqref{eq:euclidean_partition_function} reduces to a classical partition function in two dimensions. This is illustrated in Fig.~\ref{fig:2d_classical}, where we made use of the graphical notation introduced in Sec.~\ref{sec:transfer_matrix_calculations}. From Fig.~\ref{fig:2d_classical} the analogy with the field theory case displayed in Fig.~\ref{fig:2d_euclidean_QFT} becomes evident. Indeed, in the two-dimensional lattice the Euclidean time direction can now be chosen parallel to the boundaries, as shown in Fig.~\ref{fig:2d_classical_rotated_channel}. Accordingly, the partition function can be thought of as generated by iterative application of an open transfer matrix in the rotated channel, which implements the discrete time evolution.

%

The question now is: for which initial states the open transfer matrix generating the time evolution in the rotated channel is integrable? (cf. Fig.~\ref{fig:2d_classical_rotated_channel}). The answer to this question can be found as follows. By working out the algebraic steps encoded in Figs.~\ref{fig:2d_classical} and \ref{fig:2d_classical_rotated_channel} (see Ref.~\cite{PiPV17}), the computation of \eqref{eq:euclidean_partition_function} reduces to the problem of finding the leading eigenvalue of the operator
\be
\mathcal{T}(u)\sim \tensor[_{1,2}]{\langle \psi_0 |}{}T_1^{\rm QTM}(u)\otimes T_2^{\rm QTM}(-u) |\psi_0 \rangle_{1,2}\,,
\label{eq:two_row_transfer}
\ee
in $u=0$. Here we have defined 
\bea
 T_j^{\rm QTM}(u)&=&R_{2N,j}(u-w^{\ast}/2N)R_{2N-1,j}(u+w^{\ast}/2N-\eta)\nonumber\\
 &\cdots & R_{2,j}(u-w^{\ast}/2N)R_{1,j}(u+w^{\ast}/2N-\eta)\,.
\eea
A pictorial representation of the operator \eqref{eq:two_row_transfer} is given as follows :

\begin{tikzpicture}[scale=0.9]
\node at (-2.5,-.5) {\large  $\mathcal{T}(u)\sim $};
\draw[rounded corners=5pt] (0.5,-1) -- (8,-1) -- (8.5,-0.5) -- (8,0) -- (0,0) -- (-0.5,-0.5) -- (0,-1) -- (0.5,-1);
\node at (-1,-0.5) {\large $\langle \psi_0 |$};
\node at (9,-0.5) {\large $ | \psi_0\rangle$};

\foreach \x in {0.5,1.5,...,7.5} { 
\draw[] (\x,-1.25)  -- (\x,0.25);
}
\node at (0.5,-1.5) {\small $1$};
\node at (1.5,-1.5) {\small $2$};
\node at (4,-1.5) {\ldots};
\node at (6.4,-1.5) {\small $2N-1$};
\node at (7.6,-1.5) {\small $2N$};
\end{tikzpicture}

 Then, we find that the initial state is related to integrable boundaries if $\mathcal{T}(u)$ is proportional to an appropriate integrable open transfer matrix $\tau_B(u)$. Going further, one can easily infer that this can happen only if the initial state is written in terms of the matrix elements of $K(u)$ in Eq.~\eqref{eq:k_matrix_nondiag}. Indeed, following the steps sketched above (and reported in full detail in Ref.~\cite{PiPV17}), one can explicitly write down a parametrization of the family of initial states related to integrable boundaries. For example, in the XXZ spin-$1/2$ chain, this reads
\bea
 |\psi_0(\xi,\kappa,\tau)\rangle&=&- k_{12}(-\eta/2)|1,1\rangle+k_{11}(-\eta/2)|1,2\rangle\nonumber\\
 &-&k_{22}(-\eta/2)|2,1\rangle+k_{21}(-\eta/2)|2,2\rangle\,,
 \label{eq:two_site_k}
\eea
where the functions $k_{ij}(u)$ are defined in \eqref{eq:k_matrix_nondiag}. Explicitly
\bea
|\psi_0(\xi,\kappa,\tau)\rangle&=&- \kappa e^{\tau}\sinh\eta|1,1\rangle+\sinh(\xi-\eta/2)|1,2\rangle\nonumber\\
&-&\sinh(\xi+\eta/2)|2,1\rangle-\kappa e^{-\tau}\sinh\eta|2,2\rangle\,.\label{eq:boundary_state_lattice}
\eea

Eq.~\eqref{eq:boundary_state_lattice} is the final result of this
construction. Namely, we have identified a family of initial states
which play the role of the boundary states in QFT,
cf.~Sec.~\ref{sec:boundary_QFT}. In the following, we will sometimes
refer to them as \emph{lattice boundary states}. Once again, the derivation presented in this section is completely general, provided that the $R$-matrix satisfies the crossing relation \eqref{eq:crossin_relation} for an appropriate matrix $V$. Repeating the steps above, one can write down an expression for the boundary states in a generic model, which reads, up to a global numerical factor,
\be
|\psi\rangle_{1,2}=\sum_{i,j=1}^d\left[K(-\eta/2)V\right]_{i,j}|i\rangle_{1}\otimes|j\rangle_{2}\,,
\label{eq:general_boundary_state}
\ee
where we employed the notation \eqref{eq:notation_vector_basis} for the basis vectors of the local Hilbert spaces $h_j$. Here $\left[K(-\eta/2)V\right]_{i,j}$ are the matrix elements of the product $K(-\eta/2)V$, while $d$ is the dimension of the spaces $h_j$. It is straightforward to check that \eqref{eq:general_boundary_state} reduces to \eqref{eq:two_site_k} for the spin-$1/2$ chain (where $V=\sigma^{y}$), while an explicit expression for the spin-$1$ case will be provided in the following.

In the next section
we will prove that boundary states are integrable according to our definition,
as they satisfy
\eqref{eq:condition_integrability_lattice}. Incidentally, we note that
in the special case of the spin-$1/2$ chain any two-site state can
be parametrized as in \eqref{eq:boundary_state_lattice} (where one
also allows for the parameters to go to infinity). We will see
that this is not usually the case for arbitrary
models, where only a subset of two-site
product states are integrable.  

\subsection{Integrability from reflection equations}\label{sec:integrability_proof}

In the previous section we have seen that a two-site product state \eqref{eq:initial_two_site} is related to integrable boundaries in the rotated channel if its building block $|\psi_0\rangle$ is written in terms of the elements of the reflection matrix $K(u)$, see. Eq.~\eqref{eq:two_site_k}. In this section we show that for these states the condition \eqref{eq:condition_integrability_lattice} follows from general properties of integrability. As a consequence, we establish a direct connection between integrable boundaries in the rotated picture and integrability in terms of annihilation by bulk conserved charges. In turn, this unveils a direct connection with the pair structure discussed in Sec.~\ref{sec:pair_structure}.  

For the sake of clarity we once again detail the case of the XXZ spin-$1/2$ chain. However, it will be clear from our discussion that our treatment is in fact much more general, and it holds straightforwardly in all the cases where the $R$-matrix satisfies the crossing relation \eqref{eq:crossin_relation} (see, for instance, Sec.~\ref{sec:XXZ_spin-s} where the case of spin-$s$ chains is worked out). 

Consider the initial state $|\Psi_0\rangle$ defined in
\eqref{eq:initial_two_site}, where the building block $|\psi_0\rangle$
is related to integrable boundaries as in \eqref{eq:two_site_k}. We
will now prove that this state satisfies the stronger condition \eqref{eq:key_relation}, from
which \eqref{eq:condition_integrability_lattice} follows directly. We begin by defining the parameter-dependent state
\bea
|\Phi_0(u)\rangle=|\phi_0(u)\rangle_{1,2}\otimes \ldots \otimes |\phi_0(u)\rangle_{L-1,L}\,,
\label{inhomstate}
\eea
where
\bea
|\phi_0(u)\rangle&=&- k_{12}(u)|1,1\rangle+k_{11}(u)|1,2\rangle-k_{22}(u)|2,1\rangle+k_{21}(u)|2,2\rangle\,.
\label{eq:two_site_k_parameter}
\eea
From the above definition it is clear that $|\Phi_0(-\eta/2)\rangle=|\Psi_0\rangle$. Furthermore, it follows from the reflection equations \eqref{eq:reflection_equation} that
\bea
\hspace{-0.2cm}\check{R}_{3,4}(v-u)\check{R}_{2,3}(-u-v)|\phi_0(-\eta/2+u)\rangle_{1,2}\otimes|\phi_0(-\eta/2+v)\rangle_{3,4}\nonumber\\
=\check{R}_{1,2}(v-u)\check{R}_{2,3}(-u-v)|\phi_0(-\eta/2+v)\rangle_{1,2}\otimes|\phi_0(-\eta/2+u)\rangle_{3,4}\,,
\label{eq:reflection_equations_states}
\eea
where $\check{R}_{1,2}(u)=P_{1,2}R_{1,2}(u)=R_{1,2}(u)P_{1,2}$, while $P_{1,2}$ is the permutation operator
\be
P_{1,2}|v\rangle_1\otimes|w\rangle_2=|w\rangle_1\otimes|v\rangle_2\,.
\label{eq:permutation_op}
\ee
The validity of Eq.~\eqref{eq:reflection_equations_states} can be established by checking each component and making use of \eqref{eq:crossin_relation} and \eqref{eq:reflection_equation}. This is detailed for completeness in \ref{sec:technical_details}.

\begin{figure}
	\centering
	\begin{tikzpicture}[baseline={([yshift=-.5ex]current bounding box.center)}]
	\draw[line width=0.7pt,rounded corners=5pt] (-2,3) -- (-2,0.5) -- (-1.5,0) -- (-1,0.5) -- (-1,1) -- (2,2) -- (2,3);
	\draw[line width=0.7pt,rounded corners=5pt] (0,3) -- (0,0.5) -- (0.5,0) -- (1,0.5) -- (1,3);
	
\draw[-latex,line width=0.7pt]  (-2,0.7) -- (-2,0.75);	
\draw[-latex,line width=0.7pt]  (-1,0.7) -- (-1,0.75);	
\draw[-latex,line width=0.7pt]  (1,0.7) -- (1,0.75);	
\draw[-latex,line width=0.7pt]  (0,0.7) -- (0,0.75);	
	\node at (-2,3.2) {\small $1$};
	\node at (0,3.2) {\small $2$};
	\node at (1,3.2) {\small $3$};
	\node at (2,3.2) {\small $4$};
	\node at (-1.5,-0.25) {\small $-\eta/2+u$};
	\node at (0.5,-0.25) {\small $-\eta/2+v$};
	\node at (-0.5,0.8) {\scriptsize $-u-v$};
	\node at (0.6,1.2) {\scriptsize $v-u$};
	\end{tikzpicture}
	\qquad
	=
	\qquad
	\begin{tikzpicture}[baseline={([yshift=-.5ex]current bounding box.center)}]
	\draw[line width=0.7pt,rounded corners=5pt] (-2,3) -- (-2,2) -- (1,1) -- (1,0.5) -- (1.5,0.) -- (2,0.5) -- (2,3);
	\draw[line width=0.7pt,rounded corners=5pt] (-1,3) -- (-1,0.5) -- (-0.5,0) -- (0,0.5) -- (0,3);
	
	\draw[-latex,line width=0.7pt]  (2,0.7) -- (2,0.75);	
\draw[-latex,line width=0.7pt]  (-1,0.7) -- (-1,0.75);	
\draw[-latex,line width=0.7pt]  (1,0.7) -- (1,0.75);	
\draw[-latex,line width=0.7pt]  (0,0.7) -- (0,0.75);

	\node at (0.45,0.9) {\scriptsize $-u-v$};
	\node at (-0.62,1.2) {\scriptsize $v-u$};
	\node at (-2,3.2) {\small $1$};
	\node at (-1,3.2) {\small  $2$};
	\node at (0,3.2) {\small  $3$};
	\node at (2,3.2) {\small  $4$};
	\node at (-0.5,-0.25) {\small $-\eta/2+v$};
	\node at (1.5,-0.25) {\small $-\eta/2+u$};
	\end{tikzpicture}
	\caption{Pictorial representation of the reflection equations \eqref{eq:reflection_equations_states}.The orientation of the arrows reflects the fact that \eqref{eq:reflection_equations_states} is written in terms of boundary states, rather than boundary $K$-matrices.}
	\label{fig:BYBpic}
\end{figure}
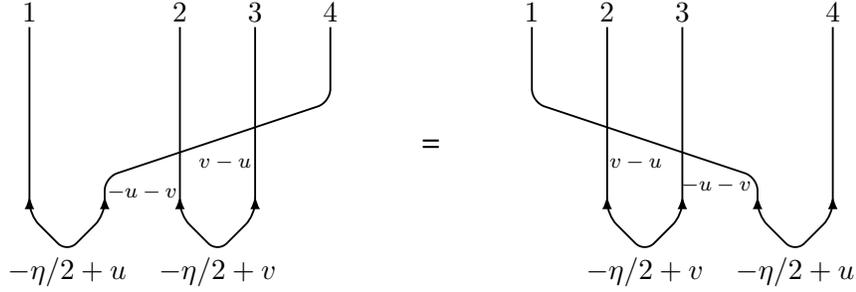

Eq.~\eqref{eq:reflection_equations_states} is a crucial relation, which is pictorially represented in Fig.~\ref{fig:BYBpic}. Even though we are now focusing on the spin-$1/2$ chain, it is in fact quite general: it is simply a rewriting of the boundary reflection equations in terms of states, rather than boundary $K$-matrices. As one should expect, for a generic model the state \eqref{eq:two_site_k_parameter} will be replaced by a different one, related to the corresponding $K$-matrix. An explicit example will be given in Sec.~\ref{sec:XXZ_spin-s} for the case of higher spin chains.

\begin{figure}
	\centering
\begin{tikzpicture}[baseline={([yshift=-.5ex]current bounding box.center)},scale=1.5]
\draw[line width=0.7pt,rounded corners=5pt] (-2,2) -- (-2,0.5) -- (-1.75,0) -- (-1.5,0.5) -- (-1.5,1) -- (3,1) -- (3,2);

\draw[-latex,line width=0.7pt]  (-2,0.7) -- (-2,0.75);	
\draw[-latex,line width=0.7pt]  (-1.5,0.7) -- (-1.5,0.75);
\foreach \x in {-1,0,...,2} 
{
	\draw[line width=0.7pt,rounded corners=5pt] (\x,2) -- (\x,0.5) -- (\x+0.25,0) -- (\x+0.5,0.5) -- (\x+0.5,2);
	\node at (\x-0.1,0.9) {\scriptsize $u$};
	\node at (\x+0.5-0.1,0.9) {\scriptsize $u$};
	\node at (\x+0.25,-0.1) {\scriptsize $-\eta/2$};
	
	\draw[-latex,line width=0.7pt]  (\x,0.7) -- (\x,0.75);	
\draw[-latex,line width=0.7pt]  (\x+0.5,0.7) -- (\x+0.5,0.75);
}
\node at (-1.75,-0.25) {\scriptsize $-\eta/2-u$};
\node at (-2,2.2) {\scriptsize $0$};
\node at (-1,2.2) {\scriptsize $1$};
\node at (-0.5,2.2) {\scriptsize $2$};
\node at (2.,2.2) {\scriptsize $L-1$};
\node at (2.5,2.2) {\scriptsize $L$};
\node at (3,2.2) {\scriptsize $L+1$};
\end{tikzpicture} 
=
\begin{tikzpicture}[baseline={([yshift=-.5ex]current bounding box.center)},scale=1.5]
\draw[line width=0.7pt,rounded corners=5pt]  (-1.5,2) -- (-1.5,1) -- (3,1) -- (3,0.5) -- (3.25,0) -- (3.5,0.5) -- (3.5,2);

\draw[-latex,line width=0.7pt]  (3,0.7) -- (3,0.75);	
\draw[-latex,line width=0.7pt]  (3+0.5,0.7) -- (3+0.5,0.75);

\foreach \x in {-1,0,...,2} 
{
	\draw[line width=0.7pt,rounded corners=5pt] (\x,2) -- (\x,0.5) -- (\x+0.25,0) -- (\x+0.5,0.5) -- (\x+0.5,2);
	\node at (\x+0.1,0.9) {\scriptsize $u$};
	\node at (\x+0.5+0.1,0.9) {\scriptsize $u$};
	\node at (\x+0.25,-0.1) {\scriptsize $-\eta/2$};
	
\draw[-latex,line width=0.7pt]  (\x,0.7) -- (\x,0.75);	
\draw[-latex,line width=0.7pt]  (\x+0.5,0.7) -- (\x+0.5,0.75);	
	
}
\node at (3.25,-0.25) {\scriptsize $-\eta/2-u$};
\node at (-1.5,2.2) {\scriptsize $0$};
\node at (-1,2.2) {\scriptsize $1$};
\node at (-0.5,2.2) {\scriptsize $2$};
\node at (2.,2.2) {\scriptsize $L-1$};
\node at (2.5,2.2) {\scriptsize $L$};
\node at (3.5,2.2) {\scriptsize $L+1$};
\end{tikzpicture} 
\caption{Graphical derivation of Eq.~\eqref{eq:subsequent_application}. The picture represents the subsequent application of the reflection equations \eqref{eq:reflection_equations_states}.}
\label{fig:BYB2}
\end{figure}
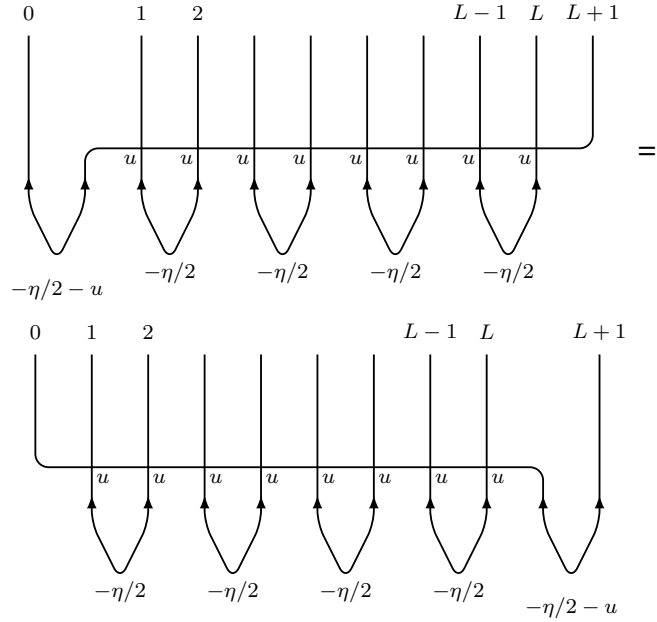

Consider now a chain of length $L$ with two auxiliary spaces $h_0$ and $h_{L+1}$, where $L$ is an even number. From repeated use of Eq.~\eqref{eq:reflection_equations_states}, as pictorially represented in Fig.~\ref{fig:BYB2}, one can prove
\bea
\hspace{-2.5cm}\check{R}_{L+1,L}(v-u)\check{R}_{L,L-1}(-v-u)\ldots \check{R}_{3,2}(v-u)\check{R}_{2,1}(-v-u)|\phi(u-\eta/2)\rangle_{0,1}\nonumber\\
\otimes |\phi(v-
\eta/2)\rangle_{2,3}\otimes \ldots
\otimes  |\phi(v-\eta/2)\rangle_{L,L+1}\nonumber\\
\hspace{-2.5cm}=\check{R}_{0,1}(v-u)\check{R}_{1,2}(-v-u)\ldots \check{R}_{L-2,L-1}(v-u)\check{R}_{L-1,L}(-v-u)|\phi(-
\eta/2+v)\rangle_{0,1}\nonumber\\
\otimes \ldots\otimes|\phi(-
\eta/2+v)\rangle_{L-2,L-1}\otimes |\phi(-\eta/2+u)\rangle_{L,L+1}\,.
\label{eq:subsequent_application}
\eea
Setting $v=0$ we get
\bea
\hspace{-2.5cm}\check{R}_{L+1,L}(-u)\ldots \check{R}_{2,1}(-u)|\phi(-\eta/2+u)\rangle_{0,1}\otimes |\phi(-
\eta/2)\rangle_{2,3}\otimes \ldots
\otimes  |\phi(-\eta/2)\rangle_{L,L+1}\nonumber\\
\hspace{-2.5cm}=\check{R}_{0,1}(-u)\ldots \check{R}_{L-1,L}(-u)|\phi(-
\eta/2)\rangle_{0,1}\otimes \ldots\otimes|\phi(-
\eta/2)\rangle_{L-2,L-1}\nonumber\\
\otimes |\phi(-\eta/2+u)\rangle_{L,L+1}\,.
\label{eq:aux_1}
\eea
From the definition of $\check{R}_{i,j}$, the l.h.s. of \eqref{eq:aux_1} can be rewritten as
\bea
\hspace{-2.5cm}R_{L+1,L}(-u)R_{L+1,L-1}(-u)\ldots R_{L+1,1}(-u)P_{L+1,L}\ldots P_{1,2}|\phi(-\eta/2+u)\rangle_{0,1}\nonumber\\
\otimes |\phi(-
\eta/2)\rangle_{2,3}\otimes \ldots
\otimes  |\phi(-\eta/2)\rangle_{L,L+1}\nonumber\\
\hspace{-2.5cm}=R_{L+1,L}(-u)R_{L+1,L-1}(-u)\ldots R_{L+1,1}(-u)P_{L+1,L}\ldots P_{1,2}P_{0,1}|W\rangle_{0,1}\otimes |\psi_0\rangle^{\otimes L/2}\,,
\label{eq:rhs}
\eea
where we introduced $|W\rangle_{i,j} =P_{i,j}|\phi(-\eta/2+u)\rangle_{i,j}$. Analogously, the r.h.s. of \eqref{eq:aux_1} yields
\bea
R_{0,1}(-u)R_{0,2}(-u)\ldots R_{0,L}(-u)P_{0,1}P_{1,2}\ldots P_{L,L+1}|\psi_0\rangle^{\otimes L/2}\otimes |W\rangle_{L,L+1}\,.
\label{eq:lhs}
\eea

\begin{figure}
	\centering
	\begin{tikzpicture}[baseline={([yshift=-.5ex]current bounding box.center)},scale=1.1]
	\draw[black!40,dashed] (0,-2.65) arc (0:180:3 and 0.3); 
	\draw[] (0,-2.65) arc (0:-180:3 and 0.3);
	
	\draw[-latex, line width=0.8pt] (-3,-2.65)+(190:3 and 0.35) arc (190:200:3 and 0.35);

	\foreach \angle in {-150,-130,...,-20}
	{ 
		\draw[rounded corners=3pt] 
		($(-3,-2.25)+(\angle-5:3 and 0.3)$) -- ($(-3,-3)+(\angle-5:3 and 0.3)$);
	}
	\node at ($(-3.,-3.25)+(-95:3 and 0.3)$) {\large $\ket{\Psi_0}$};
	\end{tikzpicture} 
	~
	=
	~
	\begin{tikzpicture}[baseline={([yshift=-.5ex]current bounding box.center)},scale=1.1]
	\draw[black!40,dashed] (0,-2.65) arc (0:180:3 and 0.3); 
	\draw[] (0,-2.65) arc (0:-180:3 and 0.3); 
	\draw[-latex, line width=0.8pt] (-3,-2.65)+(-25:3 and 0.35) arc (-25:-30:3 and 0.35); 
	\foreach \angle in {-150,-130,...,-20}
	{ 
		
		\draw[rounded corners=3pt] 
		($(-3,-2.25)+(\angle-5:3 and 0.3)$) -- ($(-3,-3)+(\angle-5:3 and 0.3)$);
	}
	\node at ($(-3.,-3.25)+(-95:3 and 0.3)$) {\large $\ket{\Psi_0}$};
	\end{tikzpicture} 
	\caption{Pictorial representation of the relation \eqref{eq:key_relation} for a generic integrable initial state $\ket{\Psi_0}$. It is proven in the main text that for states generated from boundary integrability this relation is an immediate consequence of \eqref{eq:subsequent_application}, which is depicted in Fig.~\ref{fig:BYB2}.}
	\label{fig:key_relation}
\end{figure}
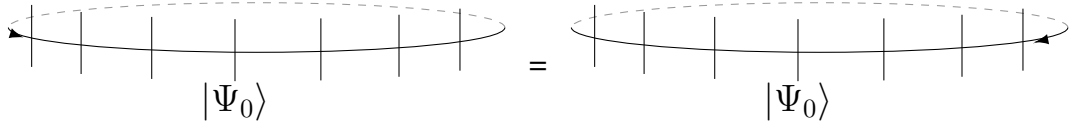

We now make use of the identities
\bea
P_{L+1,L}\ldots P_{1,2}P_{0,1}=U^{-1}\,,\\
P_{0,1}P_{1,2}\ldots P_{L,L+1}=U\,,
\eea
where $U$ is the shift operator \eqref{eq:shift_operator}. Plugging these into \eqref{eq:rhs} and \eqref{eq:lhs} we finally obtain
\bea
R_{L+1,L}(-u)R_{L+1,L-1}(-u)\ldots R_{L+1,1}(-u)|\psi_0\rangle^{\otimes L/2}\otimes |W\rangle_{0,L+1}\nonumber\\
=R_{0,1}(-u)R_{0,2}(-u)\ldots R_{0,L}(-u)|\psi_0\rangle^{\otimes L/2}\otimes |W\rangle_{0,L+1}\,.
\label{eq:almost_done}
\eea
From this equation it is now easy to conclude the proof, by writing down its components. In particular, it is shown in \ref{sec:technical_details} that \eqref{eq:almost_done} implies
\bea
{\rm tr}_0\left\{R_{L,0}(-u)R_{L-1,0}(-u)\ldots R_{1,0}(-u)\right\}|\psi_0\rangle^{\otimes L/2}\nonumber\\
={\rm tr}_0\left\{R_{0,1}(-u)R_{0,2}(-u)\ldots R_{0,L}(-u)\right\}|\psi_0\rangle^{\otimes L/2}\,,\label{eq:final_result}
\eea
which is exactly \eqref{eq:key_relation}, a pictorial representation of which is given on figure \ref{fig:key_relation}.

The proof presented in this section has far reaching
consequences. Most prominently, it directly relates boundary states on
the lattice with the pair structure frequently encountered in the
recent literature of quantum quenches. In particular, this also
provides us with a direct relation between the presence of latter and
the validity of the so-called $Y$-system. Since this is a rather
technical point, we consign its discussion to \ref{sec:y-system}.

Our proof relied on a direct application of the boundary Yang-Baxter relations, which leads to the annihilation by the odd
  charges. However, it is possible to formulate an alternative proof
  which derives the eigenvalue condition \eqref{eq:eigcond}. The idea
  is to introduce the Quantum Transfer Matrices for the inhomogeneous
  states \eqref{inhomstate}, and to use their commutativity and
  certain simple
  properties at degenerate points. For the sake of brevity
  we omit the details of this second proof.

\section{Constructing integrable matrix product states} \label{sec:MPS}

In this section we  address a systematic construction of integrable
MPSs of arbitrary bond dimension. The main idea is to obtain new
integrable MPSs starting from the known boundary two-site product
states. Once again, for the sake of clarity we focus on the XXZ
spin-$1/2$ chain, but it will be apparent that our construction is in
fact more general. 

As a first example consider the state
\be
|\chi(u_1,\ldots u_n)\rangle =\tau(u_1)\ldots \tau(u_n)|\Psi_0\rangle\,,
\label{eq:initial_MPS0}
\ee
where $|\Psi_0\rangle $ is a boundary two-site product state of the
form \eqref{eq:initial_two_site} and $\tau(u)$ is the fundamental
transfer matrix \eqref{eq:transfer}. By the proof presented in the previous section $|\Psi_0\rangle$ is integrable. Then, it follows straightforwardly from the commutativity of the transfer matrices that \eqref{eq:initial_MPS0} is also integrable, as it is annihilated by all the odd charges.

This simple observation is at the basis of our construction. More generally, further integrable states can be constructed using the so-called fused
(higher-spin) transfer matrices $\{\tau^{(d)}(u)\}_{d=1}^{\infty}$, where
we used the convention $\tau^{(1)}(u)=\tau(u)$. These operators have a similar matrix product form and
can be written as 
\be
\tau^{(d)}(u)={\rm tr}_0\left[\mathcal{L}^{(1,d)}_{L,0}(u)\ldots \mathcal{L}^{(1,d)}_{1,0}(u)\right]\,.
\label{eq:fused_transfer}
\ee
Here $\mathcal{L}^{(1,d)}_{j,0}(u)$ are the fused Lax operators, which are matrices acting on the tensor product of the local spaces $h_j\simeq \mathbb{C}^2$ and the auxiliary space $h_{0}\simeq \mathbb{C}^{d+1}$. The trace is taken over $h_0^{(d)}$, which has dimension $d+1$. In the isotropic case, $\Delta=1$, the expression for $\mathcal{L}^{(1,d)}_{1,2}(u)$ is very simple and reads \cite{Fadd96}
\be
\mathcal{L}^{(1,d)}_{1,2}(u)=u \mathbf{1} +\sum_{\alpha=1}^{3} \sigma_1^{\alpha}\otimes S^{\alpha}_2\,,
\label{eq:fused_l-matrix}
\ee
where $\mathbf{1}$ is the identity operator on the space $h_1\otimes
h_2$, $\sigma^{\alpha}$ are the Pauli matrices while $S^{\alpha}$ are
the operators corresponding to the standard $(d+1)$-dimensional
representation of $SU(2)$. In the anisotropic case, $\Delta\neq 1$,
Eq.~\eqref{eq:fused_l-matrix} has to be replaced by an appropriate
deformed expression, involving the generators of the quantum group
$U_{q}(sl_2)$ \cite{grs-96}. The operators \eqref{eq:fused_transfer}
are called fused transfer matrices, as they can be obtained from
\eqref{eq:transfer} by an appropriate procedure named fusion
\cite{KRS-81}. 

Importantly, all the transfer matrices \eqref{eq:fused_transfer} commute,
\be
\left[\tau^{(d_1)}(u),\tau^{(d_2)}(v)\right]=0\,,
\label{eq:fused_commutation}
\ee
so that we can immediately construct an infinite family of integrable MPSs. Consider
\be
|\chi_{d_1,\ldots , d_{n}}(u_1,\ldots u_n)\rangle =\tau^{(d_1)}(u_1)\ldots \tau^{(d_n)}(u_n)|\Psi_0\rangle\,,
\label{eq:initial_MPS}
\ee
where $|\Psi_0\rangle $ is an arbitrary boundary two-site product state of the
form \eqref{eq:initial_two_site}. It follows  immediately, that \eqref{eq:initial_MPS} is integrable, as
\be
Q_{2r+1}|\chi_{d_1,\ldots , d_{n}}(u_1,\ldots u_j)\rangle=\tau^{(d_1)}(u_1)\ldots \tau^{(d_j)}(u_j)Q_{2r+1}|\Psi_0\rangle=0\,,
\ee
where we used that the fused transfer matrices commute with the local
charges (which follows from \eqref{eq:def_charges_norm} and
\eqref{eq:fused_commutation}). The state \eqref{eq:initial_MPS} can be cast into the canonical MPS form 
\bea
\hspace{-2cm}|\chi_{d_1,\ldots , d_{n}}(u_1,\ldots u_n)\rangle = \sum_{i_1,\dots,i_L=1}^D{\rm tr}_{\mathcal{H}_{D}}\left[A^{(i_1)}(u_1,\ldots u_n)B^{(i_2)}(u_1,\ldots u_n)A^{(i_3)}(u_1,\ldots u_n)\right.\nonumber\\
\left. B^{(i_4)}(u_1,\ldots u_n) \ldots  A^{(i_{L-1})}(u_1,\ldots u_n)B^{(i_{L})}(u_1,\ldots u_n) \right]|i_1,\ldots ,i_L\rangle\,,
\label{eq:canonical_MPS_form_integrable}
\eea
where $\mathcal{H}_D$ is an auxiliary space of dimension
$D=\prod_{j=1}^n(d_j+1)$. Here $A^{(i)}_j(u_1,\ldots u_n)$,
$B^{(i)}_j(u_1,\ldots u_n)$ are such that the product
$A^{(i)}_j(u_1,\ldots u_n)B^{(i)}_j(u_1,\ldots u_n)$ is a $D\times D$
dimensional matrix and their explicit form can be derived from the
knowledge of the operators $\mathcal{L}^{(1,d_j)}(u_j)$. Note that
$|\chi\rangle$ in \eqref{eq:canonical_MPS_form_integrable} is in
general two-site invariant, analogously to the state
$|\Psi_0\rangle$.

\begin{figure}
	\begin{center}
		\begin{tikzpicture}
		\foreach \x in {1,2.5,4}{
			\draw (\x, -2) -- (\x, 4);
		}
		\draw[red!80!black!,line width =1.5] (5.5,-2)  -- (5.5,4);
		\foreach \x in {-1.5,-0.5,0.5,...,3.5}{
			\draw (0.5,\x) -- (6, \x);
			\draw[dashed] (0,\x) -- (0.5,\x);
		};
		\foreach \x in {-1.5,0.5,...,2.5}{
			\draw[rounded corners=2pt] (6,\x) -- (6.25,\x+0.25)  -- (6.5,\x+0.5) --  
			(6.25, \x+0.75) -- (6,\x+1);
			\node at (7,\x+0.5) {$|\psi_0\rangle $};
		}
		\foreach \x in {1,2.5,4}{
			\draw[dashed] (\x,-2.5) -- (\x,-2);
			\draw[dashed] (\x,4) -- (\x,4.5);
		};
		\draw[red!80!black!,line width =1.5,dashed] (5.5,-2.5) -- (5.5,-2);
		\draw[red!80!black!,line width =1.5,dashed] (5.5,4) -- (5.5,4.5);
		\draw[-latex, line width =2, draw=blue!50!white] (8,-2.5) -- (8,4.5);
		\node at (8.1, 4.7) {$t=\infty$};
		\node at (8.3, -2.8) {$t=-\infty$};
		
		\end{tikzpicture}
		\caption{Pictorial representation of the quench from an integrable MPS in the rotated channel. By choosing the Euclidean time direction along the boundary, in analogy with Fig.~\ref{fig:2d_euclidean_QFT}, the initial state is interpreted as a boundary in space. The Euclidean time evolution can then be thought of as generated by an open transfer matrix in which the reflection matrix is specified by the initial state. If the latter is of the form \eqref{eq:initial_MPS}, the corresponding open transfer matrix is integrable.}
		\label{fig:2d_classical_rotated_channel_with_MPS}
	\end{center}
\end{figure}
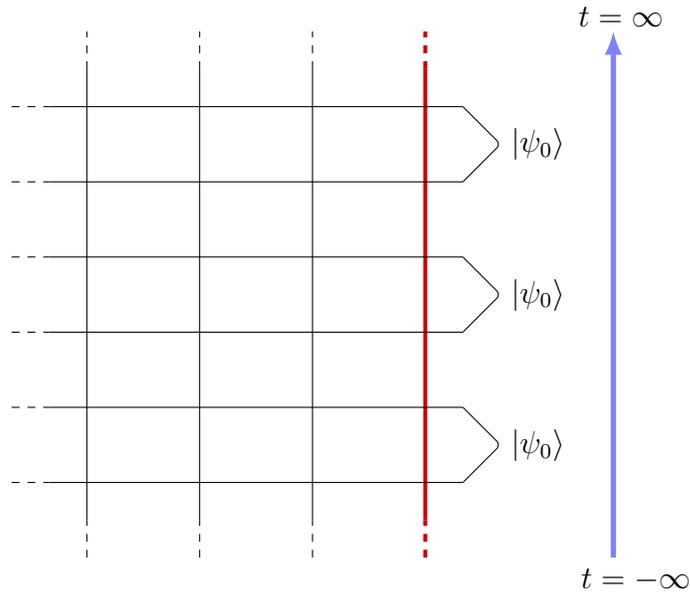

We note that the MPSs \eqref{eq:initial_MPS} could be interpreted as lattice versions of the ``smeared boundary states'' introduced in the study of quantum quenches in the context of conformal field theories \cite{cc-06,Card16-2,Card17-RG} Importantly, the MPSs \eqref{eq:initial_MPS} can also be
related to integrable boundaries in the rotated channel. Consider for example the state 
\be
|\chi_d(w)\rangle=\tau^{(d)}(w)|\Psi_0\rangle\,,
\ee
where $|\Psi_0\rangle$ is defined in \eqref{eq:initial_two_site}, with its building block $|\psi_0\rangle$ satisfying \eqref{eq:two_site_k}. Then, in analogy with \eqref{eq:representation_transfer} one has the following pictorial representation 
\begin{equation}
|\chi_d \rangle =
\begin{tikzpicture}[baseline=(current  bounding  box.center),scale=1.5]
  \draw[ red!60!black!20, dashed,line width=2pt] (0,-2.65) arc (0:180:3 and 0.3); 
 \draw[red!80!black!,line width=2pt] (0,-2.65) arc (0:-180:3 and 0.3);
 
 \draw[red!80!black!,-latex, line width=2.5pt] (-3,-2.65)+(190:3 and 0.35) arc (190:200:3 and 0.35);

\foreach \angle in {-150,-130,...,-20}
{ 

\draw[rounded corners=4pt] ($(-3,-2.75)+(\angle:3 and 0.3)$)  -- ($(-3,-2)+(\angle:3 and 0.3)$);
\draw[rounded corners=3pt] 
($(-3,-2)+(\angle-10:3 and 0.3)$) -- ($(-3,-3)+(\angle-10:3 and 0.3)$)  --  ($(-3,-3.25)+(\angle-5:3 and 0.3)$)  -- ($(-3,-3)+(\angle:3 and 0.3)$)  -- ($(-3,-2.75)+(\angle:3 and 0.3)$) ;

\node at ($(-3.,-3.5)+(\angle-5:3 and 0.3)$) {\small $\ket{\psi_0}$};
}
\label{eq:representation_integrable}
\end{tikzpicture}
\end{equation}
Here we indicated the auxiliary row (corresponding to a Hilbert space
of dimension $d+1$) with a thick red line, to distinguish it from the
$2$-dimensional representation appearing in the usual transfer matrix
\eqref{eq:transfer}. One can consequently repeat the steps outlined in
Sec.~\ref{sec:integrable_boundaries}: in this case the time evolution
in the rotated channel is represented pictorially in
Fig.~\ref{fig:2d_classical_rotated_channel_with_MPS}. We see that
application of $\tau^{(d)}(w)$ only results in the insertion of a line
in the two-dimensional partition function with respect to the
situation displayed in Fig.~\ref{fig:2d_classical_rotated_channel}. In
particular, using the properties of fused transfer matrices one can
see that the open transfer matrix appearing in this construction is
still integrable. Finally, it is clear that the same happens by
applying a finite number of transfer matrices $\tau^{(d_j)}(w_j)$, as
this amounts to the insertion of a finite number of legs in the
corresponding open transfer matrix. 

From \eqref{eq:initial_MPS}, we see that the set of integrable states
is infinite, and their construction involves a large number of free
parameters. Even fixing the number $n$ of transfer matrices, their
bond dimensions $d_j$ and spectral parameters can still be chosen
arbitrarily. It is an important question whether this family exhausts
all possibilities of integrable MPSs with finite bond
dimension. While we can not give a definite answer to this
question, it is remarkable that all known cases indeed fit into this
framework, including the MPSs studied in
\cite{LeKZ15,BLKZ16}. This will be detailed in the
following section.

An alternative  way to understand our construction is to interpret the MPSs as non-scalar solutions
to the boundary Yang-Baxter equations. Then the task of
finding all integrable MPSs can be split into two parts: determining
whether the boundary Yang-Baxter equations are necessary for the integrability condition
to hold, and finding all finite dimensional solutions in terms of
local Lax operators acting on local $K$-matrices. We leave this problem for future research.

To conclude this section we note that the
overlaps between integrable MPSs of the form \eqref{eq:initial_MPS} and the eigenstates of the
Hamiltonian are immediately obtained if the overlaps with the state
$|\Psi_0\rangle$ are known. This follows from the fact that the
transfer matrices act diagonally on the Bethe states and the
eigenvalues are known from the algebraic Bethe ansatz, see for example \cite{Suzu99} for explicit formulas. Employing the same notation of Sec.~\ref{sec:pair_structure}, we have
\be
\langle\{\lambda_j\}|\prod_{r=1}^k\tau^{(d_r)}(w_r)|\Psi_0\rangle=\left[\prod_{r=1}^k\tau^{(d_r)}(w_r|\{\lambda_j\})\right] \langle\{\lambda_j\} |\Psi_0\rangle\,,
\ee
where $\tau^{(d_r)}(w_r|\{\lambda_j\})$ is the eigenvalue of $\tau^{(d_r)}(w_r)$ corresponding to the eigenstate $|\{\lambda_j\}\rangle$.

\section{Integrable quenches: analysis of specific models}\label{sec:integrable_qunches}

In this last section we  review several recent studies of
quantum quenches in different models, where closed-form analytical
results could be obtained. We show that in all of these cases the initial states are integrable according to our
definition. These include the MPSs constructed in the works \cite{LeKZ15,BLKZ16}, which are shown to fit into the framework of the previous sections.
We also present new results by producing concrete
formulas for the integrable two-site states of the spin-1 XXZ chain.

\subsection{The XXZ spin-$1/2$ chain}\label{sec:xxz_case}

We begin our analysis with the prototypical case of the XXZ spin-$1/2$ chain \eqref{eq:hamiltonian_xxz}, where quench problems have been extensively investigated in the past few years \cite{PiPV17,Pozs13,gritsev-stb-neel-to-xx-first,gritsev-demler-stb-quench-osszf-neel-to-xx,MoCa10,LiAn14,WDBF14,MPTW15,PiVC16, 
  FaEs13,FCEC14,IDWC15,DeCD16,PVCR17}.
In particular, a large number of
      analytical results have been obtained. Exact overlap formulas
      between assigned initial states and the eigenstates of
      \eqref{eq:hamiltonian_xxz} have been derived for special classes
      of two-site states
      \cite{KoPo12,Pozs14,BNWC14,PiCa14,BDWC14-2,Broc14} and, in the
      isotropic case, for more general matrix product
      states\cite{LeKZ15,BLKZ16,FoZa16}. Closed-form results for
      the long-time limit of local observables were derived in
      \cite{WDBF14,MPTW15}, by means of the Quench Action method
      \cite{CaEs13,Caux16}, while exact computations for the
      Loschmidt echo have been reported in \cite{Pozs13,PiPV17} for
      arbitrary two-site product states. Finally, an exact result for
      the time evolution of entanglement entropies has been obtained
      in \cite{AlCa17}.

From the previous section, it is now clear that all two-site states
considered in these works are integrable, as they are boundary
states. These include, as a special case, the N\'eel and the so-called
dimer states studied in \cite{WDBF14,MPTW15}, but also the tilted
N\'eel and ferromagnet states considered in \cite{PiVC16}. Note that
even though these states are very simple, the computation of their
overlaps is extremely difficult: in fact, the latter are still unknown
in the case of tilted N\'eel and ferromagnet states. Nevertheless, it
follows from our derivation that for generic $\Delta$ the pair
structure holds for all local two-site states. 

Among the most interesting recent developments was the discovery of
exact overlap formulas for MPSs in the isotropic case $\Delta=1$ \cite{LeKZ15,BLKZ16,FoZa16}. The overlaps were shown to have the same
structure as in the case of the N\'eel state: they included the same Gaudin-like
determinants, and only the pre-factors were different. Here we show how these MPSs can be embedded
into our framework of integrable initial states. In particular, we
argue that they can be obtained by the action of (fundamental or fused)
transfer matrices on simple two-site states; in the first few examples
we explicitly calculate the corresponding dimensions and spectral
parameters.

In \cite{BLKZ16} the following family of states was considered 
\bea
|\chi^{(k)}_0\rangle &=& {\rm tr}\left[\prod_{l=1}^L\left(S_x^{(k)}|1\rangle_l+S_y^{(k)}|
2\rangle_l \right)\right]\nonumber\\
&=&\sum_{i_1,\dots,i_L=1}^d{\rm tr}\left[Z^{(i_1)}\ldots Z^{(i_L)}\right]|i_1,\ldots , i_L\rangle \,,
\label{eq:zarembo_MPS}
\eea
where
\bea
Z^{(1)}&=&S^{(k)}_x\,,\\
Z^{(2)}&=&S^{(k)}_y\,.
\eea
Here $S^{(k)}_\alpha$ are the $(k+1)\times (k+1)$ matrices corresponding to the standard representation of $SU(2)$ generators, and the trace in \eqref{eq:zarembo_MPS} is over the associated $(k+1)$-dimensional space. 

In order to test integrability of these MPSs we numerically constructed
the corresponding QTMs \eqref{eq:monodromy_MPS} in the first few
cases. By evaluating the eigenvalue condition
\eqref{eq:eigcond} we confirmed integrability of these
states. In fact, this also directly follows from the
  corresponding overlap formulas which were computed in \cite{BLKZ16},
  explicitly displaying the pair structure.

 The question of how these states fit into our
  framework of boundary integrability is not immediately clear from
  their MPS representation.
  Below we show
  by explicit calculations that the two
simplest vectors $|\chi_0^{(1)}\rangle$ and
  $|\chi_0^{(2)}\rangle$ are the translationally invariant components of
  simple two-site product states; we believe that this is a new
  result. Going further, the structure of the next few states
  could be investigated using a recursion relation derived in
   \cite{BLKZ16}, 
   which expresses
  $|\chi_0^{(k)}\rangle$ in terms of $|\chi_0^{(k-2)}\rangle$ and
  $|\chi_0^{(k-4)}\rangle$. This relation explicitly involves
  fundamental transfer matrices of the form \eqref{eq:transfer} with
  an auxiliary space of dimension $2$ and is such that higher MPSs
   are expressed as sums of lower ones. Our goal is to relate these to integrable boundaries, namely we intend to express higher
  MPSs in the product form \eqref{eq:initial_MPS}. This could be
  achieved by a careful analysis of the recursion formulas of
  \cite{BLKZ16} and by the so-called
  $T$-system, a set of functional relations for different fused transfer
  matrices \cite{Suzu99}. However, the full implementation of this program is beyond
  the scope of this paper, and we content ourselves with deriving
  explicit formulas for $|\chi_0^{(k)}\rangle$ for the first few values of $k$.

The model is invariant under the global action of $SU(2)$, so we are allowed to perform
global rotations of the physical space (if necessary, this rotation
can easily be restored at the end of our calculations). We consider the operator 
\be
\mathcal{W}=W_1\otimes \dots \otimes W_L\,,
\label{eq:global_rotation}
\ee
where $W_j$ is a local rotation at the site $j$, described by the matrix
\be
W=\frac{e^{i\pi/4}}{\sqrt{2}}
\left(
\begin{array}{cc}
	1 & i\\
	1 & -i
\end{array}
\right)\,.
\ee
After applying the operator \eqref{eq:global_rotation}, the state \eqref{eq:zarembo_MPS} is rewritten, up to an irrelevant global numerical phase, as
\bea
|\chi^{(k)}_0\rangle &=&\sum_{i_1,\dots,i_L=1}^d{\rm tr}\left[\tilde{Z}^{(i_1)}\ldots \tilde{Z}^{(i_L)}\right]|i_1,\ldots , i_L\rangle \,,
\label{eq:psiK_state}
\eea
where now
\bea
\tilde{Z}^{(1)}&=&S^{(k)}_+\equiv \frac{1}{\sqrt{2}}\left[ S^{(k)}_x+iS^{(k)}_y\right]\,,\label{eq:tilde_zeta_1}\\
\tilde{Z}^{(2)}&=&S^{(k)}_-\equiv \frac{1}{\sqrt{2}}\left[ S^{(k)}_x-iS^{(k)}_y\right]\,.
\label{eq:tilde_zeta_2}
\eea

For $k=1$, it is immediate to simplify this expression. In this case $(S^{\pm})^2=0$, and we get immediately
\be
\label{eq:nedec}
|\chi_0^{(1)}\rangle \propto |N\rangle + U|N\rangle\,,
\ee
where $U$ is the shift operator \eqref{eq:shift_operator}, while $|N\rangle$ is the N\'eel state,
\be
|N\rangle = |1\rangle_1 \otimes |2\rangle_2 \otimes  \ldots \otimes |1\rangle_{L-1} \otimes |2\rangle_L\,. 
\label{eq:neel_state}
\ee
We see that, for $k=1$, \eqref{eq:psiK_state} is nothing but the
zero-momentum component of the N\'eel state. Restoring the rotation
by \eqref{eq:global_rotation} we see that in this simplest case the
MPS \eqref{eq:zarembo_MPS} is the zero-momentum component of the
tilted N\'eel state in the $y$-direction:
\be
\label{eq:nedec2}
\hspace{-2cm} |\chi_0^{(1)}\rangle \propto (1+U)|N_y\rangle,\qquad 
|N_y\rangle\equiv\left[\frac{1}{2}\left(|1,1\rangle+|2,2\rangle+i|1,2\rangle-i|2,1\rangle  \right)\right]^{\otimes L/2}
\,.
\ee

A decomposition similar to \eqref{eq:nedec} can be 
 performed for generic $k$, using standard
techniques in the literature of MPSs. 
In particular, one can prove that \eqref{eq:psiK_state} is always the
zero-momentum component of a $2$-periodic MPS. The proof of this
statement is reported in \ref{sec:MPS_states}, where we derive the
following decomposition 
\be
|\chi^{(k)}_0\rangle= |\Phi^{(k)}_0\rangle+ U|\Phi^{(k)}_0\rangle\,,
\label{eq:decomposition_zero_momentum}
\ee
where 
\bea
|\Phi^{(k)}_0\rangle &=&\sum_{i_1,\dots,i_L=1}^d{\rm tr}\left\{A^{(i_1)}B^{(i_2)}A^{(i_3)}B^{(i_4)}\ldots A^{(i_{N-1})}B^{(i_N)} \right\}|i_1,\ldots ,i_N\rangle\,.
\eea
Here we have defined
\bea
A^{(1)} &= &\mathcal{P}_1S^{+}\mathcal{P}_2\,,\\
A^{(2)} &= &\mathcal{P}_1S^{-}\mathcal{P}_2\,,\\
B^{(1)} &= &\mathcal{P}_2S^{+}\mathcal{P}_1\,,\\
B^{(2)} &= &\mathcal{P}_2S^{-}\mathcal{P}_1\,,
\eea
where we introduced the projectors $\mathcal{P}_1$ and $\mathcal{P}_2$, which are diagonal matrices with elements
\begin{equation}
\label{projectors}
(\mathcal{P}_1)_{j,j}=\frac{1+(-1)^j}{2}\,, \qquad   (\mathcal{P}_2)_{j,j}=\frac{1-(-1)^j}{2}\,.
\end{equation}
One can now check for each value of $k$ that $|\Phi^{(k)}\rangle$ is
of the form \eqref{eq:initial_MPS} for an appropriate choice of the
two-site state $|\psi_{0}\rangle$. We have done this explicitly up to
$k=5$. In particular, we obtained
\bea
|\Phi_0^{(2)}\rangle&\propto&\left(|1,2\rangle+|2,1\rangle\right)^{\otimes L/2}\,,\\
|\Phi_0^{(3)}\rangle&\propto&\tau^{(1)}(1) U|N\rangle\,,\\
|\Phi_0^{(4)}\rangle&\propto&\tau^{(1)}(3/2)\left(|1,2\rangle+|2,1\rangle\right)^{\otimes L/2}\,,\\
|\Phi_0^{(5)}\rangle&\propto&\tau^{(2)}(3/2)U|N\rangle\,,
\eea
where $|N\rangle$ is the N\'eel state \eqref{eq:neel_state} while $U$
is the shift operator. After rotating back with the inverse of
\eqref{eq:global_rotation}, we obtain the following list for the
original MPSs:
\bea
|\chi_0^{(2)}\rangle&\propto&(1+U)\left(|1,1\rangle+|2,2\rangle\right)^{\otimes L/2}\,,\\
|\chi_0^{(3)}\rangle&\propto&(1+U)\tau^{(1)}(1) |N_y\rangle\,,\\
|\chi_0^{(4)}\rangle&\propto&(1+U)\tau^{(1)}(3/2)\left(|1,1\rangle+|2,2\rangle\right)^{\otimes L/2}\,,\\
|\chi_0^{(5)}\rangle&\propto&(1+U)\tau^{(2)}(3/2)|N_y\rangle\,.
\eea
This representation is a new result of our
work. We conjecture that all $|\chi_0^{(k)}\rangle$ can be
written in a similar product form using higher spin fused transfer
matrices. The proof of this conjecture and the derivation of explicit formulas is beyond the scope of this work.

As a final remark on the spin-$1/2$ XXZ model, we note that
additional integrable MPSs can be constructed for special 
values of the anisotropy $\Delta$, in the regime $\Delta<1$. These
correspond to the so-called ``root of unity points", where additional
representations of the underlying quantum group, and hence of transfer
matrices, exist \cite{grs-96}. MPSs obtained using these additional transfer matrices can be incorporated
into our discussion, because they still commute
with the transfer matrix~\eqref{eq:transfer}.

\subsection{The XX model}\label{sec:xx_model}

At the non-interacting point $\Delta=0$ the XXZ Hamiltonian \eqref{eq:hamiltonian_xxz} deserves special attention. In this case, one recovers the so-called XX chain, which has served as a prototypical benchmark for countless studies on non-equilibrium dynamics in isolated many-body systems. In particular, analytic results for global quantum
quenches have been presented in
\cite{gritsev-stb-neel-to-xx-first,gritsev-demler-stb-quench-osszf-neel-to-xx,XX-quench-brockmann}, where the dynamics arising from the N\'eel state was mainly addressed.
This model presents an important example where integrable states do not display the pair structure discussed in Sec~\ref{sec:pair_structure}. In particular, the special structure of conserved charges lead to a different constraint for the rapidities of the eigenstates annihilated by the odd ones. This is briefly discussed in the following, while we refer to  \cite{XX-quench-brockmann} for a systematic treatment of quantum quenches from the N\'eel state. 

We recall that the XX model can be studied
by
introducing fermionic operators through the Jordan-Wigner transformation
\begin{equation*}
c_j^\dagger=e^{i\pi \sum_{k=1}^{j-1} \sigma_k^+\sigma_k^-}  \sigma_j^+\qquad
c_j=e^{-i\pi \sum_{k=1}^{j-1} \sigma_k^+\sigma_k^-} \sigma_j^-\,,
\end{equation*}
which satisfy the relations:
\begin{equation}
\label{fermcomm}
\{c_j,c_k\}=  \{c_j^\dagger,c_k^\dagger\}=0,\qquad
\{c_j^\dagger,c_k\}=\delta_{j,k}\,.
\end{equation}
The Hamiltonian is then written as $H=\sum_p  \eps_p  \tilde c_p^\dagger \tilde
c_p$, where
\begin{equation*}
\tilde c_p^\dagger=\frac{1}{\sqrt{L}}\sum_{j=1}^L e^{ipj} c_p^\dagger\,,\qquad
\tilde c_p=\frac{1}{\sqrt{L}}\sum_{j=1}^L e^{-ipj} c_j\,,
\end{equation*}
and $\eps_p=\cos(p)$.

Alternatively, the model can be studied by considering the limit $\Delta\to 0$ (namely,
$\eta\to i\pi/2$) of the Bethe ansatz solution of \eqref{eq:hamiltonian_xxz} for generic $\Delta$. In the
latter language the quasi-particles are associated to the rapidities
$\lambda$, which are related to the lattice momentum through
\begin{equation}
\label{rapiid}
e^{ip}=\frac{\sinh(\lambda+i\pi/4)}{\sinh(\lambda-i\pi/4)}\,.
\end{equation}
It follows from the Bethe equations that the Bethe rapidities are either real or they have imaginary part equal to $\pi/2$, and the corresponding one-particle energy eigenvalues are
\begin{equation}
e(\lambda)=-\frac{1}{\cosh(2\lambda)}\,.
\end{equation}

As in the case of generic $\Delta$, a set of conserved charges $Q_j$ can be generated via the transfer matrix construction explained in Sec.~\ref{sec:general_setting}. Alternatively, a set of charges can be constructed using the
fermionic operators, such that their commutativity follows from \eqref{fermcomm}.
The relation between the two sets of charges is non-trivial and has
been studied in detail in \cite{GM-higher-conserved-XXZ}, see also \cite{Fago14,FaEs13-2}. From the fermionic point of view, one possible choice for the charges is simply the set of occupation numbers
$\tilde n_k$ for the Fourier modes. These operators are inherently
non-local, but they form a complete commuting family. An
alternative choice is to consider the local operators
\begin{equation*}
I_j=\sum_{k}  e^{ijk} \tilde n_k=\sum_{l=1}^L   c^\dagger_l c_{l+j-1}\,,
\end{equation*}
and the Hermitian combinations
\begin{equation}
\label{Ipm}
I^+_j=I_j+I_{-j}\,,  \qquad     I^-_j=\frac{(I_j-I_{-j})}{i}\,.
\end{equation}
It follows directly from the commutation relations \eqref{fermcomm} that every
$I_j$ is conserved. The charges $I^+_j$ are even and the $I^-_j$ are
odd under space reflection.

On the other hand, it was shown in
\cite{GM-higher-conserved-XXZ} that in the XX model all charges can be
expressed using the operators
\begin{equation}
e_n^{\alpha\beta}=\sum_{j=1}^L e_{n,j}^{\alpha\beta},\qquad
e_{n,j}^{\alpha\beta}=\sigma_j^\alpha\sigma_{j+1}^z \dots
\sigma_{j+n-2}^z \sigma_{j+n-1}^\beta\,.
\end{equation}
In particular, there are two families of charges defined as
\begin{equation}
H^+_n=
\cases{e_n^{xx}+e_n^{yy}\,, & $n$  even\,,\\
	e_n^{xy}-e_n^{yx}\,, & $n$  odd\,,}
\end{equation}
and
\begin{equation}
H^-_n=
\cases{e_n^{xy}-e_n^{yx}\,,      & $n$  even\,,\\
	e_n^{xx}+e_n^{yy}\,,  & $n$ odd\,.}
\end{equation}
The canonical charges $Q_j$ obtained from the transfer matrix are linear
combinations of the first family. Explicit formulas can be found in
\cite{GM-higher-conserved-XXZ}: in the simplest example $Q_3=-H^+_3$, for which the one-particle eigenvalues read
\begin{equation}
q_3(\lambda)=\frac{2\sinh(2\lambda)}{\cosh^2(2\lambda)}\,.
\end{equation} 
Using the Jordan-Wigner transformation it can be seen that the
families $\{H_n^{\pm}\}$ and $\{I_n^\pm\}$ contain the same operators, but with
alternating identification:
\begin{equation}
H^+_n=\cases{
	2I^+_n\,,  & $n$ even,\\
	-2I^-_n\,, & $n$  odd,}\qquad \qquad
H^-_n=\cases{
	-2I^-_n\,,  & $n$ even,\\
	2I^+_n\,, & $n$  odd.}
\end{equation}
In the present case, integrability of the initial state is equivalent to
annihilation by the charges $H^+_n$ for all odd $n$. In the following, we show that this does not imply the pair structure.

Consider the rapidity transformation
\begin{equation}
\label{raptraf}
\lambda \to i\pi/2-\lambda\,.
\label{eq:rapidity_transformation}
\end{equation}
It is straightforward to verify that this implies
\begin{equation}
e(\lambda)\to -e(\lambda)\,, \qquad q_3(\lambda)\to q_3(\lambda)\,.
\end{equation}
Analogously, one can see that the eigenvalues of all odd charges
are invariant under \eqref{eq:rapidity_transformation}. Taking an arbitrary eigenstate, this
transformation can be performed for every rapidity individually, so that a large number of states can be generated which share the
same eigenvalues for all odd $Q_j$.  Eigenstates corresponding to sets of rapidities \eqref{eq:pair_structure} are
obviously annihilated by the odd charges, but so do all the other
states obtained after repeated application of \eqref{eq:rapidity_transformation}. The rapidities of these eigenstates do not uniquely display pairs of opposite rapidities. Since integrable states will in general overlap with all eigenstates annihilated by odd conserved charges, it follows from this discussion that they will not posses the pair structure. This is explicitly shown in the case of the N\'eel state in \cite{XX-quench-brockmann}, to which we refer for further detail.

As a final remark, we note that the XX chain is closely related to the $q\to\infty$
limit of the $q$-boson model studied in
\cite{sajat-q2,sajat-qboson}. However, the two models are connected by
a highly non-local transformation, which alters the locality of the
charges. A detailed analysis of the latter for the $q$-boson model is
out of the scope of the present paper. 


\subsection{XXZ chains with higher spin}\label{sec:XXZ_spin-s}

Analytic results for quantum quenches in higher-spin chains were presented in \cite{PiVC16}. In particular, closed-form characterizations of post-quench steady states were obtained for the spin-$1$ chain known as the Zamolodchikov-Fateev model \cite{ZaFa80}. The Hamiltonian reads
\bea
H_{\rm ZF}^{(1)}&=&\sum_{j=1}^{L}\Big\{\left[S^{x}_jS^{x}_{j+1}+S^{y}_jS^{y}_{j+1}+\cosh\left(2\eta\right) S^{z}_jS^{z}_{j+1}\right]\nonumber\\
&+&2\left[(S^{x}_{j})^{2}+(S^{y}_{j})^{2}+\cosh\left(2\eta\right)(S^{z}_{j})^{2}\right] \nonumber\\
&-&\sum_{a,b}A_{ab}(\eta)S^{a}_jS^{b}_jS^{a}_{j+1}S^{b}_{j+1}\Big\},
\label{eq:hamiltonian_spin_1}
\eea 
where the indices $a,b$ in the second sum take the values $x$, $y$, $z$ and where periodic boundary conditions are assumed, $S^{\alpha}_{N+1}=S^{\alpha}_1$. The coefficients $A_{ab}$ are defined by $A_{ab}(\eta)=A_{ba}(\eta)$ and
\bea
A_{xx} & = & A_{yy} = A_{xy}= 1\,,\\
A_{zz} & = & \cosh\left(2\eta\right)\,,\\
A_{xz} & = & A_{yz}=2\cosh\eta-1\,,
\label{eq:coefficients}
\eea
while $\eta$ plays the role of the anisotropy parameter along the $z$-direction. Finally the operators $S^{\alpha}_j$ are given by the standard three-dimensional representation of the $SU(2)$ generators
\bea
\fl S^x=\frac{1}{\sqrt{2}}\left(\begin{array}{c c c}0&1&0\\1&0&1\\0&1&0\end{array}\right),\quad S^y=\frac{1}{\sqrt{2}}\left(\begin{array}{c c c}0&-i&0\\i&0&-i\\0&i&0\end{array}\right),\quad S^z=\left(\begin{array}{c c c}1&0&0\\0&0&0\\0&0&-1\end{array}\right).
\label{eq:spin_op}
\eea

Two particular initial states were considered in \cite{PiVC16}, namely
\bea
|\Psi_1\rangle=|1,3\rangle_{1,2} \otimes \ldots  \otimes|1,3\rangle_{L-1,L}\,,\label{eq:s1}\\
|\Psi_2\rangle=|2,2\rangle_{1,2}\otimes \ldots  \otimes|2,2\rangle_{L-1,L}\,,\label{eq:s2}
\eea
where $|j\rangle$, $j=1,2,3$ represent the basis vectors of the local Hilbert space $h\simeq \mathbb{C}^3$. For these states analytical formulas were obtained by means of a $Y$-system, cf. \ref{sec:y-system}. Not surprisingly, one can show that these states are integrable according to our definition. In particular, they belong to the class of boundary states, as we detail in the following.

The $R$-matrix corresponding to the Hamiltonian \eqref{eq:hamiltonian_spin_1} reads \cite{InOZ96}
\be
R_{12}(u)=\frac{1}{\sinh(u+2\eta)}
\left(
\begin{array}{ccccccccc}
	a_1&&&&&&&&      \\
	&a_2&&a_3&&&&&   \\
	&&a_4&&a_5&&a_6&&\\
	&a_3&&a_2&&&&&   \\
	&&a_5&&a_7&&a_5&&\\
	&&&&&a_2&&a_3&   \\
	&&a_6&&a_5&&a_4&&\\
	&&&&&a_3&&a_2&   \\
	&&&&&&&&a_1
\end{array}
\right)\,,
\label{eq:R_matrix_spin_1}
\ee
where
\bea
	a_1=\sinh(u+2\eta)\,,\qquad a_2=\sinh u\,,\\
	a_3=\sinh 2\eta\,,\qquad a_4 = \frac{\sinh u\sinh(u-\eta)}{\sinh(u+\eta)}\,,\\
	a_5=\frac{\sinh 2\eta\sinh u}{\sinh(u+\eta)}\,, \qquad a_6=\frac{\sinh\eta \sinh 2\eta}{\sinh(u+\eta)}\,,\\ 
	a_7=a_6+\sinh u\,.
\eea
As usual, the normalization of the $R$-matrix is chosen such that the corresponding transfer matrix both satisfies the regularity condition \eqref{eq:regularity_conditions} and has charges $Q_n$ with the correct even/odd behavior \eqref{eq:parity_charges}. One can verify that the $R$-matrix satisfies \eqref{eq:crossin_relation} with
\be
V=
\left(
\begin{array}{ccc}
	0& 0 &1 \\
	0 & -1& 0\\
	1 & 0 & 0
\end{array}
\right)\,,
\label{eq:gauge_spin_1}
\ee
and $\gamma(u)=\sinh(-u)\sinh(-u+\eta)/(\sinh(u+\eta)\sinh(u+2\eta))$. As a consequence, one can parallel the derivation outlined in Sec.~\ref{sec:integrable_boundaries} and identify a class of lattice boundary states, analogously to \eqref{eq:boundary_state_lattice}. In order to do this, one also needs to introduce the $K$-matrix \cite{InOZ96}
\be
K(u)=
\left(
\begin{array}{ccc}
	x_1(u)&y_1(u)&z(u)\\
	\tilde{y}_1(u)&x_2(u)&y_2(u)\\
	\tilde{z}(u)&\tilde{y}_2(u)&x_3(u)
\end{array}
\right),
\label{eq:K:spin_1}
\ee
where
\bea
y_1(u)&=&\mu\sinh\left(\zeta-\frac{\eta}{2}+u\right)\sinh 2u\,,\\
y_2(u)&=&\mu\sinh\left(\zeta+\frac{\eta}{2}-u\right)\sinh 2u\,,\\
z(u)&=&\mu^2\frac{\sinh\left(\eta-2u\right)}{2\cosh\eta}\sinh 2u\,,\\
x_1(u)&=&
\sinh\left(\frac{\eta}{2}+\zeta+u\right)\sinh\left(\frac{\eta}{2}-\zeta-u\right)\nonumber\\
&+&\mu\tilde{\mu}\frac{\sinh(\eta-2u)}{2\cosh\eta}\sinh\eta\,, 
\\
x_2(u)&=&
\sinh\left(\frac{\eta}{2}+\zeta-u\right)\sinh\left(\frac{\eta}{2}-\zeta-u\right)\,,
\nonumber\\
&+&\mu\tilde{\mu}\frac{\sinh(\eta-2u)}{2\cosh\eta}\sinh(\eta+2u)\,, \\
x_3(u)&=&
\sinh\left(\frac{\eta}{2}+\zeta-u\right)\sinh\left(\frac{\eta}{2}-\zeta+u\right)\nonumber\\
&+&\mu\tilde{\mu}\frac{\sinh\left(\eta-2u\right)}{2\cosh\eta}\sinh\eta\,. 
\eea
Here $\zeta,\mu,\tilde{\mu},$ are arbitrary constants, while functions $\tilde{y}_1(u)$, $\tilde{y}_2(u)$ and $\tilde{z}(u)$ are defined analogously with the substitution $\mu\to\tilde{\mu}$.

The calculations sketched in Sec.~\ref{sec:integrable_boundaries} can be repeated straightforwardly for the spin-$1$ case. Indeed, since the $R$-matrix satisfies the crossing relation, the general formula \eqref{eq:general_boundary_state} applies. Then, plugging in \eqref{eq:general_boundary_state} the explicit expression for the gauge matrix \eqref{eq:gauge_spin_1} and the reflection matrix \eqref{eq:K:spin_1}, one obtains the following class of boundary states  
\bea
|\Psi(\mu,\tilde{\mu},\zeta)\rangle =|\psi(\mu,\tilde{\mu},\zeta)\rangle_{1,2}\otimes \ldots |\psi(\mu,\tilde{\mu},\zeta)\rangle_{L-1,L}\,,
\label{eq:boundary_states_spin_1}
\eea
where
\bea
\fl \hspace{0.5cm}|\psi(\mu,\tilde{\mu},\zeta)\rangle = -\mu ^2 \sinh ^2(\eta ) |1,1\rangle + \mu  \sinh (\eta ) \sinh (\zeta -\eta )|1,2\rangle \nonumber\\
\fl \hspace{1cm}+\left[\mu  \tilde{\mu} \sinh ^2(\eta )-\sinh (\zeta ) \sinh (\zeta -\eta )\right]|1,3\rangle 
- \mu  \sinh (\eta ) \sinh (\zeta +\eta )|2,1\rangle \nonumber\\
\fl\hspace{1cm} +\sinh (\zeta -\eta ) \sinh (\zeta +\eta )|2,2\rangle-\tilde{\mu } \sinh (\eta ) \sinh (\zeta -\eta )|2,3\rangle \nonumber\\
\fl \hspace{1cm}+ \left[\mu  \tilde{\mu } \sinh ^2(\eta )-\sinh (\zeta ) \sinh (\zeta +\eta )\right]|3,1\rangle+\tilde{\mu } \sinh (\eta ) \sinh (\zeta +\eta )|3,2\rangle\nonumber\\
\fl \hspace{1cm} -\tilde{\mu }^2 \sinh ^2(\eta )|3,3\rangle\,.
\label{eq:boundary_states_spin_1_block}
\eea
It is easy to check that the states \eqref{eq:s1} and
\eqref{eq:s2} belong to this class. Then, from the proof of
Sec.~\ref{sec:integrable_boundaries}, which also applies for the
spin-$1$ case, we immediately obtain that they satisfy the condition
\eqref{eq:condition_integrability_lattice}, and hence they are
integrable. Note that the integrability of the class
\eqref{eq:boundary_states_spin_1} can also be verified by constructing
the Quantum Transfer Matrix \eqref{eq:monodromy_MPS}, and numerically
evaluating the corresponding eigenvalues
in a neighborhood of $u=0$.

Contrary to the spin-$1/2$ formula \eqref{eq:two_site_k_parameter}, it
is not true that all two-site states can be parametrized as in
\eqref{eq:boundary_states_spin_1_block}. In other words, boundary
states for the spin-$1$ model are only a subset of the two-site
states: one can explicitly check that for arbitrary choices of the
latter \eqref{eq:integrability_test_thermodynamic} does not hold. This
is actually true for all the spin-$s$ generalizations of the XXZ
Hamiltonian \eqref{eq:hamiltonian_xxz} with $s\geq 1$. These models
can be obtained by the well-known fusion procedure \cite{KRS-81}, from
which also the corresponding $K$-matrices can be built
\cite{Zhou96}. As the spin $s$ increases, the dimension of the local
Hilbert spaces $h_j$ will also increase. On the other hand,  there are always just $3$ free parameters of
the fused $K$-matrices,
and they are not enough to parametrize all the
states in the tensor product $h_j\otimes h_{j+1}$.

\subsection{The $SU(3)$-invariant chain}

\label{sec:su3}

Finally, we touch upon higher rank generalization of the $SU(2)$ chain, and focus on the $SU(3)$-invariant Lai-Sutherland Hamiltonian \cite{Lai74}
\be
H_L=\sum_{j=1}^{L}\left[{\bf S}_j\cdot {\bf S}_{j+1}+\left({\bf S}_j\cdot {\bf S}_{j+1}\right)^2\right]\,.
\label{eq:hamiltonian_SU(3)}
\ee
Here the local Hilbert space is $h_j\simeq \mathbb{C}^3$, while $S^{\alpha}_j$ are given once again by the standard three-dimensional representation of the $SU(2)$ generators \eqref{eq:spin_op}. The analytical description of this model is significantly more involved as it requires a nested Bethe ansatz treatment \cite{efgk-05}. Nevertheless, many elements of the algebraic construction discussed in section \ref{sec:general_setting} are also valid in this case. The $R$-matrix of the model is
\be	
R_{1,2}(\lambda)=\frac{\lambda}{\lambda+i} + \frac{i}{\lambda+i} P_{1,2},
\label{eq:r_matrx}
\ee
where $P_{1,2}$ is the permutation matrix \eqref{eq:permutation_op}. The transfer matrix can be simply obtained by \eqref{eq:transfer}.

Recently, a remarkable overlap formula was conjectured in \cite{LeKM16} for a particular matrix product state, with a form which is reminiscent of the one in $SU(2)$ chains. The state is
\bea
\ket{\Psi_0} &=& \,{\rm tr}_0\left[\prod_{j=1}^L\Big(\sigma_0^x\ket{1}_j+\sigma^y_0\ket{2}_j+\sigma^z_0\ket{3}_j\Big)\right]\nonumber\\
&=& \sum_{\{\alpha_j\}}{\rm tr}_0\left[\sigma^{\alpha_1}\sigma^{\alpha_2}\ldots \sigma^{\alpha_L}\right]\ket{e_{\alpha_1},e_{\alpha_2},\ldots ,e_{\alpha_L}}\,,
\label{eq:initial_state}
\eea
where the trace is over the auxiliary space $h_0\simeq \mathbb{C}^2$,
and where $\sigma^{\alpha}_0$ are Pauli matrices acting on $h_0$. Note
that the auxiliary space has a different dimension from the physical
spaces $h_j\simeq \mathbb{C}^3$. This state is translational
invariant, and the integrability conditions
can easily be tested. We
constructed the corresponding QTM 
and verified the eigenvalue condition \eqref{eq:eigcond}  numerically, demonstrating that
the state \eqref{eq:initial_state} is indeed integrable. Note that this also follows
from the conjectured form of the overlap with the Bethe states
\cite{LeKM16}, from which the pair structure is evident.  

An important question is whether this state can be understood in terms
of integrable boundary conditions in the rotated channel. On the one
hand, in the $SU(3)$ case the study of integrable boundary transfer
matrices is more complicated \cite{AACD04}. On the other hand, in the
argument of Sec.~\ref{sec:integrability_proof} we explicitly used the
crossing relation \eqref{eq:crossin_relation} of the $R$-matrix, which
is no longer true for the $SU(3)$ model \cite{DoNe98}. Accordingly, care has to be
taken to generalize those constructions to this case. We hope to
return to these topics
in a future work.

\section{Conclusions}\label{sec:conclusions}

In this work we have proposed a definition of integrable states for
quantum quenches in lattice integrable systems, which is directly
inspired by the classical work on boundary quantum field theory of
Ghoshal and Zamolodchikov \cite{GhZa94}. We have identified integrable
states as those which are annihilated by all the odd local conserved
charges of the Hamiltonian. We have proven that these include the states which can be
related to integrable boundaries in an appropriate rotated channel, in
loose analogy with QFT. Furthermore, we have shown that in  all of the known
cases where closed-form analytical results could be obtained, the
initial state is integrable according to our definition. 

In the prototypical case of XXZ spin-$s$ chains we have shown that integrable
states include two-site product states together with larger families
of MPSs.  In the spin-1/2 chain all two-site states are integrable,
whereas for higher spin this is true only for a subset of them. We
have characterized this subset in terms of the fused $K$-matrices. 

One of the properties of integrable quenches seemed to be the pair structure
for the overlaps, because this was observed in simple cases in the
spin-1/2 XXZ chain and also the Lieb-Liniger model
\cite{DWBC14,BNWC14,Broc14,WDBF14,MPTW15}.
The pair structure has important consequences for the entropy of the
	steady state arising after a quantum quench \cite{AlCa17-QR,PVCR17,BeTC17}, therefore it is important to
clarify its relation with the integrability of the initial state.
We have argued that the
pair structure indeed follows from integrability for generic values of the 
coupling constants (and could actually be proven for the isotropic
Heisenberg chain).
Together with our proof of integrability
this constitutes a general confirmation of the pair structure for a
wide variety of states, including already known cases and new states
where the actual overlaps are not yet known.
On the other hand, we have also discussed possible exceptions to the
pair structure, such as the
XX 
model detailed in Sec. \ref{sec:xx_model}. Nevertheless, the integrability condition
\eqref{eq:condition_integrability_lattice} can be always introduced
without modifications.

Remarkably, in almost all of the cases encountered, MPSs annihilated by the odd
charges could be understood as boundary states in the rotated
channel. It is an important open question whether this family exhausts all
integrable MPSs with finite bond dimensions.
In the $SU(3)$-invariant model the MPS introduced in \cite{LeKM16} and
studied here in \ref{sec:su3} is integrable according to our
definition, but its interpretation in terms of boundary integrability
is not known yet. We hope to return to this question in future work.

As a final remark, we stress that our results should not be
interpreted as a no-go 
theorem for obtaining exact results for non-integrable initial
states. However, our work provides a unified point of view for the
many exact results that appeared in the past years. Furthermore, if
exact results for time evolution are to be derived  (be it results 
for correlators or the Loschmidt echo or other quantities), 
 one should first look at the integrable initial states, regardless of the model
considered. The test of integrability is straightforward,
therefore our framework gives an extremely useful starting point for the study
of models where quantum quenches have not yet been investigated. An
example is the case of $SU(N)$-invariant spin chains, where only
a small number of results are currently available \cite{LeKM16,MBPC17}. 

\section{Acknowledgments}
We are very grateful to Pasquale Calabrese and G\'{a}bor
Tak\'{a}cs for inspiring discussions and useful comments.  B. P. is
grateful to Pasquale Calabrese and SISSA for their hospitality.
E. V. acknowledges
support by the ERC under Starting Grant 279391
EDEQS. B. P. acknowledges support from the ``Premium'' Postdoctoral
Program of the Hungarian Academy of Sciences, and the K2016 grant no. 119204 of the research agency NKFIH.

\appendix

\section{Integrability condition for $p$-periodic states}\label{sec:proof_integrability_condition}

In this appendix we show that the validity of \eqref{eq:integrability_test} implies the annihilation of all odd conserved charges, namely \eqref{eq:condition_integrability_lattice}.
For simplicity, we assume that the initial state is two-site shift invariant with $\langle\Psi_0|\Psi_0\rangle=1$, as an analogous derivation holds in the general case.

First, it follows from \eqref{eq:formal_representation} that one can write down the following formal expansion 
\begin{equation}
G(u)= \bra{\Psi_0}
\exp\left[2\sum_{k=1}^\infty \alpha_{2k}\frac{u^{2k}}{(2k)!} Q_{2k+1}\right]
\ket{\Psi_0}\,.
\end{equation}
Note that here we used $U^2\ket{\Psi_0}=\ket{\Psi_0}$ and that for parity invariant states $G(u)=\tilde G(u)$. At the first orders in $u^2$ we have
\bea
\fl G(u)=1+2\frac{u^2}{2} \alpha_2\bra{\Psi_0}Q_3\ket{\Psi_0}+
4\frac{u^4}{4}\alpha_2^2\bra{\Psi_0}Q_3Q_3\ket{\Psi_0}+2\frac{u^4}{4!}\alpha_4\bra{\Psi_0}Q_5\ket{\Psi_0}+\dots
\eea
If $G(u)\equiv 1$ then we have immediately that
$\bra{\Psi_0}Q_3\ket{\Psi_0}=0$. However, at the next order we have a
sum of two terms. If the state is parity invariant, then
$\bra{\Psi_0}Q_5\ket{\Psi_0}=0$ and we also have 
\begin{equation}
\label{q3v}
\bra{\Psi_0}Q_3Q_3\ket{\Psi_0}=0\,.
\end{equation}
If the state is not parity invariant, then we use the information coming from $\tilde
G(u)$. Using that the latter is also identically $1$, we get the vanishing of
the two terms separately.

If \eqref{q3v} holds then $Q_3\ket{\Psi_0}=0$, namely
this means that $Q_3$ can be neglected from the Taylor series
completely. Going to the next order, we get the vanishing of the mean value
of $Q_7$, and similarly we can prove
$Q_5\ket{\Psi_0}=0$. Proceeding iteratively, the annihilation by all odd charges is proven.

\section{Numerical study of the Quantum Transfer Matrix for integrable
  and non-integrable MPSs}\label{sec:examples_QTM_construction}

\begin{figure}
	\centering
	\begin{subfigure}
		\centering
		\includegraphics[scale=0.65]{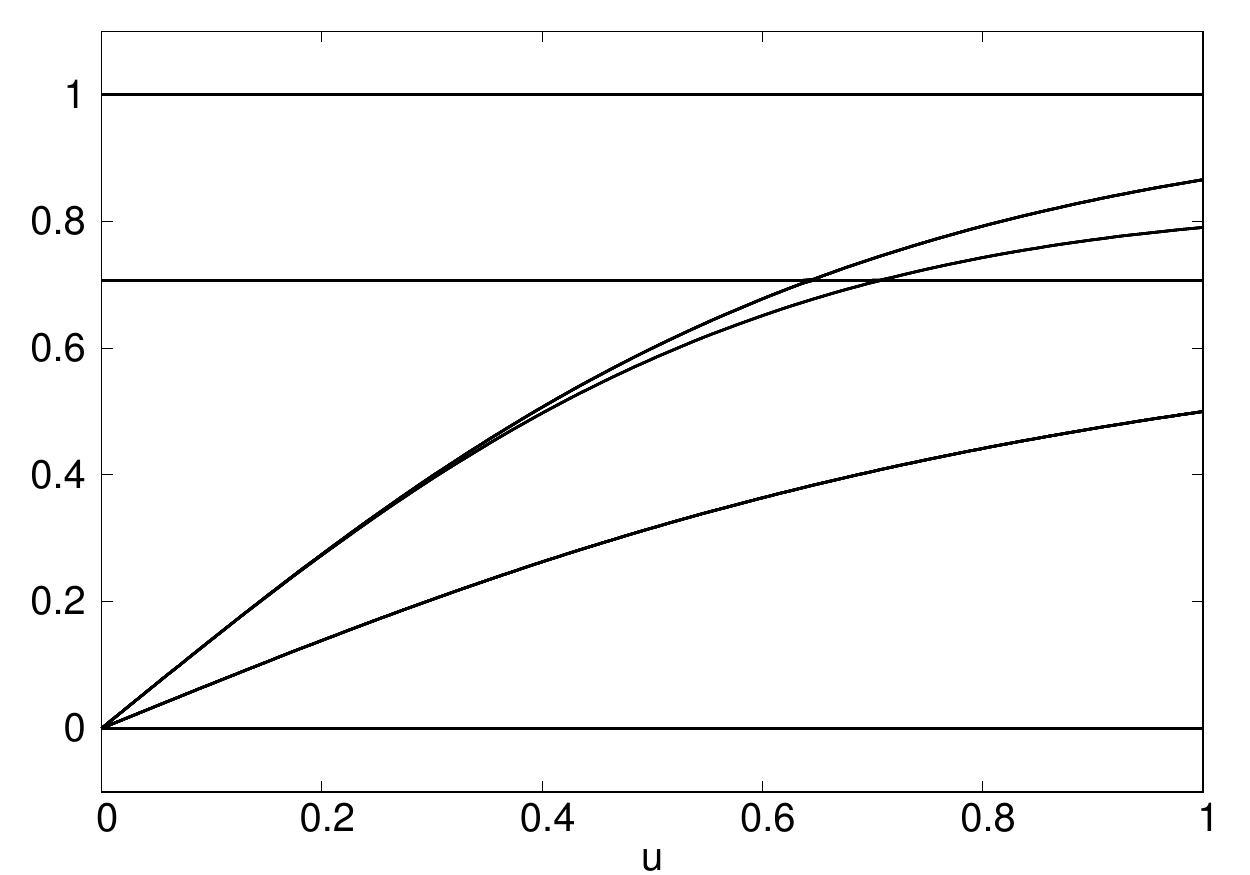}
		\caption{Magnitude of the eigenvalues of the operator
                  $\tilde{\mathcal{T}}(u)$ corresponding to the
                  (translationally invariant) dimer
                  state \eqref{eq:dimer_state}. We see that the
                  eigenvalues which are non-zero for $u=0$ remain
                  constant for a non-vanishing neighborhood of
                  $u=0$. In fact, they are completely $u$-independent,
                but level crossings can occur. The eigenvalues with the
                second largest magnitude describe the exponentially
                decaying overlap between the original and the one-site
                shifted dimer states.}
		\label{E1}
	\end{subfigure}
	\begin{subfigure}
		\centering
		\includegraphics[scale=0.65]{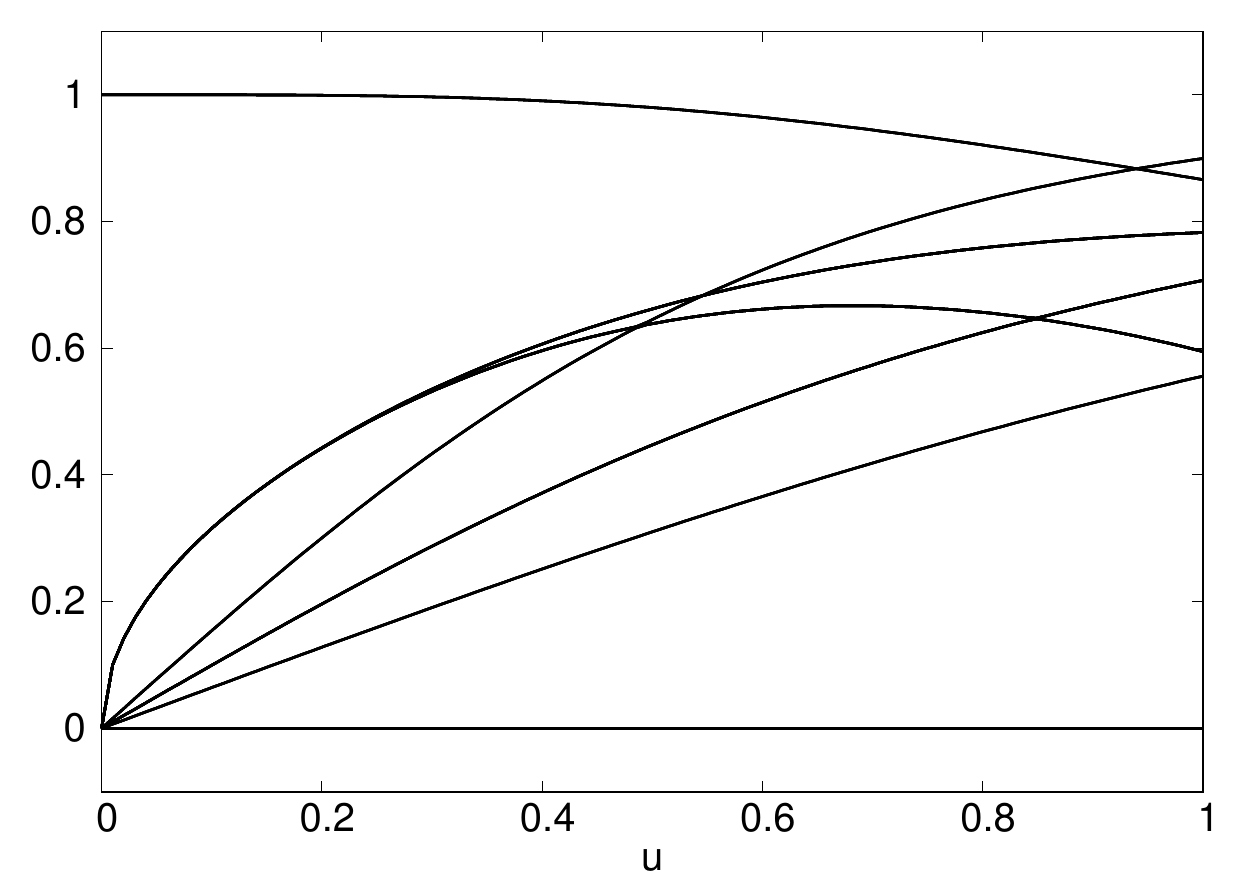}
		\caption{Magnitude of the eigenvalues of the operator $\tilde{\mathcal{T}}(u)$ corresponding to the four-site state \eqref{eq:four_site}. One can see numerically that there is no finite neighborhood of $u=0$ such that the eigenvalues which are non-zero for $u=0$ remain constant.}
		\label{E2}
	\end{subfigure}
\end{figure}

In this appendix we numerically study the matrix $\tilde \mathcal{T}(u)$ introduced in
\eqref{eq:monodromy_MPS} and compute its eigenvalues for both an
integrable and a non-integrable case.

For the integrable case, we consider the zero-momentum dimer state
defined as
\begin{equation}
  \ket{\Psi_0}\sim \left(1+U\right) \times \left[\otimes_{j=1}^{L/2}   \frac{\ket{1,2}-\ket{2,1}}{\sqrt{2}}\right]\,.
  \label{eq:dimer_state}
\end{equation}
It is easy to check that this state is generated by a
translationally invariant MPS of the form \eqref{eq:translationally_invariant_MPS}, with
\begin{equation}
  A^{(1)}=
  \left(
\begin{array}{ccc}
         0 & 1 & 0\\
   0 & 0 & 0\\
   1/\sqrt{2} & 0 & 0\\
\end{array}
\right)\,,
 \qquad
  A^{(2)}=
 \left(
\begin{array}{ccc}
   0 & 0 & 1\\
   -1/\sqrt{2}  & 0 & 0\\
  0 & 0 & 0\\
  \end{array}
\right)\,.
\end{equation}
As a non-integrable state, we choose the translationally invariant
component of a four-site domain-wall state, namely 
  \begin{equation}
  \ket{\Psi_0}\sim \left(1+U+U^2+U^3\right) \times \left[\otimes_{j=1}^{L/4}   \ket{1,1,2,2}\right]\,.
  \label{eq:four_site}
\end{equation}
This state was already considered in \cite{PiVC16} (and mentioned in
\cite{IQNB16}), where it was shown that the corresponding rapidity distribution functions did not satisfy the $Y$-system, cf.~\ref{sec:y-system}. Therefore it is natural
to expect that this state is not integrable according to our definition. The state in \eqref{eq:four_site} can still be written as a translational invariant MPS of the form \eqref{eq:translationally_invariant_MPS}, where now
\begin{equation}
    A^{(1)}=
  \left(
\begin{array}{cccc}
  0 & 1 & 0 & 0 \\
  0 & 0 & 1 & 0 \\
  0 & 0 & 0 & 0 \\
  0 & 0 & 0 & 0 \\
\end{array}
\right)\,,
 \qquad
  A^{(2)}=
 \left(
\begin{array}{cccc}
  0 & 0 & 0 & 0 \\
  0 & 0 & 0 & 0 \\
  0 & 0 & 0 & 1 \\
  1 & 0 & 0 & 0 \\
\end{array}
\right)\,.
\end{equation}

We computed the eigenvalues of the QTM $\tilde{\mathcal{T}}(u)$ for these two MPSs. For concreteness, we focused on the XXX chain, with the $R$-matrix
normalization given by $R(u)=(u+iP)/(u+i)$ (here $P$ is the permutation operator defined in \eqref{eq:permutation_op}). The QTM is symmetric with respect to
the sign of $u$ and in the following we restricted to positive values of $u$. The numerical
results for the magnitude of the eigenvalues are
shown in Figs.~\ref{E1} and \ref{E2}. Note that a single curve here
corresponds to at least two different eigenvalues due to the sign
difference. Further degeneracies can be present, but we can omit to specify them as they are irrelevant to the test of integrability. 

In the case of the dimer state, we see from Fig.~\ref{E1} that the eigenvalues of $\tilde{\mathcal{T}}(u)$ which are
non-zero for $u=0$ remain constant for a non-vanishing neighborhood of $u=0$. On the contrary, we see from Fig.~\ref{E2} that this is not true for the case of the state \eqref{eq:four_site}. We conclude that the QTM method confirms
integrability of the dimer state and its breaking for the four-site domain-wall state \eqref{eq:four_site}.

\section{Integrability of boundary states: technical details}\label{sec:technical_details}

In this appendix we provide further technical details on the proof presented in Sec.~\ref{sec:integrability_proof}. We start by showing that \eqref{eq:reflection_equations_states} is a simple rewriting of the reflection equations \eqref{eq:reflection_equation}. 

Given the operator $V$ acting on $h_1$, we use the following convention for its matrix elements
\be
\langle i|V|j\rangle=V_{i,j}=V^i_j\,,
\ee
so that
\be
V|j\rangle=V^{i}_j|i\rangle\,.
\ee
Analogously, given the operator $R_{1,2}(u)$ acting on $h_1\otimes h_2$, we use 
\be
\tensor[_1]{\langle i|}{}\tensor[_2]{\langle j|}{}R_{1,2}(u)|l\rangle_1|k\rangle_2=R_{lk}^{ij}(u)\,,
\ee
so that
\be
R_{1,2}(u)|l, k\rangle = R_{lk}^{ij}(u)|i,j\rangle\,.
\ee
Using these conventions, the matrix elements of the partial transposed are easily written
\be
\tensor[_1]{\langle i|}{}\tensor[_2]{\langle j|}{}R^{t_1}_{1,2}(u)|l\rangle_1|k\rangle_2=R_{ik}^{lj}(u)\,.
\ee

Consider now the matrix $\Gamma(u)=iK(u)V$, where $K(u)$ is the reflection matrix \eqref{eq:k_matrix_nondiag}, while $V$ is introduced in \eqref{eq:crossin_relation}. One can verify that the vector $\eqref{eq:two_site_k_parameter}$ can be written as
\be
|\phi(u)\rangle_{1,2}=\Gamma_{i,j}(u)|i\rangle_{1}\otimes|j\rangle_{2}=\Gamma_{j}^i(u)|i\rangle_{1}\otimes|j\rangle_{2}\,.
\ee
Using the formula above it is a simple exercise to rewrite \eqref{eq:reflection_equations_states} as
\bea
\hspace{-1.2cm}\Gamma^{j_1}_{i_4}(-\eta/2+u)\Gamma^{i_2}_{i_3}(-\eta/2+v)R^{j_2i_1}_{i_2i_4}(-u-v)R^{j_3j_4}_{i_3i_1}(v-u)|j_1,j_2,j_3,j_4\rangle\nonumber\\
\hspace{-0.8cm}=\Gamma^{i_1}_{i_2}(-\eta/2+v)\Gamma^{i_3}_{j_4}(-\eta/2+u)R^{i_4j_3}_{i_3i_2}(-u-v)R^{j_1j_2}_{i_4i_1}(v-u)|j_1,j_2,j_3,j_4\rangle\,,
\eea
from which it follows
\bea
\hspace{-1.2cm}\Gamma^{j_1}_{i_4}(-\eta/2+u)\Gamma^{i_2}_{i_3}(-\eta/2+v)R^{j_2i_1}_{i_2i_4}(-u-v)R^{j_3j_4}_{i_3i_1}(v-u)\nonumber\\
\hspace{-0.8cm}=\Gamma^{i_1}_{i_2}(-\eta/2+v)\Gamma^{i_3}_{j_4}(-\eta/2+u)R^{i_4j_3}_{i_3i_2}(-u-v)R^{j_1j_2}_{i_4i_1}(v-u)\,.
\eea
Using
\be
R^{\alpha\beta}_{\gamma\delta}(u)=R^{\beta\alpha}_{\delta\gamma}(u)\,,
\ee
it is now straightforward to verify that the above expression can be rewritten as
\bea
\hspace{-1.6cm}\tensor[_1]{\langle j_2|}{}\tensor[_2]{\langle j_1|}{}\Gamma_2(-\eta/2+u)R^{t_2}_{1,2}(-u-v)\Gamma_1(-\eta/2+v)R_{1,2}^{t_1t_2}(v-u)|j_3\rangle_1|j_4\rangle_2\nonumber\\
\hspace{-1.2cm}=\tensor[_1]{\langle j_2|}{}\tensor[_2]{\langle j_1|}{}R_{1,2}(v-u)\Gamma_1(-\eta/2+v)R^{t_1}_{1,2}(-u-v)\Gamma_2(-\eta/2+u)|j_3\rangle_1|j_4\rangle_2\,.
\eea
The two matrices above have the same matrix elements in the basis $|i\rangle_1\otimes |j\rangle_2$ of $h_1\otimes h_2$ and thus are equal. Using now the definition of $\Gamma$ and the crossing relation \eqref{eq:crossin_relation} one finally obtains
\bea
K_2\left(-\eta/2+u\right)R_{1,2}(u+v-\eta)K_{1}\left(-\eta/2+v\right)R_{1,2}(v-u)\nonumber\\
=R_{1,2}(v-u)K_{1}\left(-\eta/2+v\right)R_{1,2}(u+v-\eta)K_2\left(-\eta/2+u\right)\,,
\eea
which is equivalent to the reflection equation \eqref{eq:reflection_equation}. 

Analogous algebraic manipulations can be done to derive \eqref{eq:final_result} starting from \eqref{eq:almost_done} as we now detail. First we observe that
\be
\hspace{-1cm}|W\rangle_{0,L+1}=P_{0,L+1}\Gamma^i_{j}(-\eta/2+u)|i\rangle_0|j\rangle_{L+1}=\Gamma^j_{i}(-\eta/2+u)|i\rangle_0|j\rangle_{L+1}\,.
\ee
Next, we rewrite \eqref{eq:almost_done} in components as
\bea
\hspace{-1.6cm}\Gamma^{i_1}_{i_2}(-\eta/2)\Gamma^{i_3}_{i_4}(-\eta/2)\ldots \Gamma^{i_{L-1}}_{i_L}(-\eta/2)R^{\alpha_1 j_1}_{i_{L+1}i_1}(-u)R^{\alpha_2 j_2}_{\alpha_1 i_2}(-u)R^{\alpha_3 j_3}_{\alpha_2 i_3}(-u)\ldots\nonumber\\
\hspace{-1.4cm} R^{j_{L+1} j_L}_{\alpha_{L-1},i_L}(-u)\Gamma^{i_{L+1}}_{j_0}(-\eta/2+u)|j_0,j_1,\ldots j_{L+1}\rangle\nonumber\\
\hspace{-1.2cm}=\Gamma^{i_1}_{i_2}(-\eta/2)\Gamma^{i_3}_{i_4}(-\eta/2)\ldots \Gamma^{i_{L-1}}_{i_L}(-\eta/2)R^{\alpha_1 j_L}_{i_{0}i_L}(-u)R^{\alpha_2 j_{L-1}}_{\alpha_1 i_{L-1}}(-u)R^{\alpha_3 j_{L-2}}_{\alpha_2 i_{L-2}}(-u)\ldots \nonumber\\
\hspace{-0.6cm}R^{j_{0} j_1}_{\alpha_{L-1},i_1}(-u)\Gamma^{j_{L+1}}_{i_0}(-\eta/2+u)|j_0,j_1,\ldots j_{L+1}\rangle\,.
\eea
Since this holds for every choice of $j_0$ and $j_{L+1}$, it implies
\bea
\hspace{-1.6cm}\Gamma^{i_1}_{i_2}(-\eta/2)\Gamma^{i_3}_{i_4}(-\eta/2)\ldots \Gamma^{i_{L-1}}_{i_L}(-\eta/2)R^{\alpha_1 j_1}_{i_{L+1}i_1}(-u)R^{\alpha_2 j_2}_{\alpha_1 i_2}(-u)R^{\alpha_3 j_3}_{\alpha_2 i_3}(-u)\ldots\nonumber\\
\hspace{-1.4cm} R^{j_{L+1} j_L}_{\alpha_{L-1},i_L}(-u)\Gamma^{i_{L+1}}_{j_0}(-\eta/2+u)|j_1\rangle_1\otimes \ldots \otimes |j_{L}\rangle_{L}\nonumber\\
\hspace{-1.2cm}=\Gamma^{i_1}_{i_2}(-\eta/2)\Gamma^{i_3}_{i_4}(-\eta/2)\ldots \Gamma^{i_{L-1}}_{i_L}(-\eta/2)R^{\alpha_1 j_L}_{i_{0}i_L}(-u)R^{\alpha_2 j_{L-1}}_{\alpha_1 i_{L-1}}(-u)R^{\alpha_3 j_{L-2}}_{\alpha_2 i_{L-2}}(-u)\ldots \nonumber\\
\hspace{-0.6cm}R^{j_{0} j_1}_{\alpha_{L-1},i_1}(-u)\Gamma^{j_{L+1}}_{i_0}(-\eta/2+u)|j_1\rangle_1\otimes \ldots \otimes |j_{L}\rangle_{L}\,.
\eea
Once again, it is straightforward to verify that the above expression can be rewritten as
\bea
\hspace{-1.8cm}\tensor[_0]{\langle j_0|}{}\Gamma^{t_0}_0(-\eta/2+u) R^{t_0}_{0,1}(-u)R^{t_0}_{0,2}(-u)\ldots R^{t_0}_{0,L}(-u)|j_{L+1}\rangle_{0} |\Psi_0\rangle_{1,2,\ldots L}=\nonumber\\
\hspace{-1.4cm}\tensor[_0]{\langle j_0|}{}R_{0,1}(-u)R_{0,2}(-u)\ldots R_{0,L}(-u)\Gamma^{t_0}_0(-\eta/2+u)|j_{L+1}\rangle_{0}  |\Psi_0\rangle_{1,2,\ldots L}\,,
\eea
where
\be
|\Psi_0\rangle_{1,2,\ldots L}=|\phi(-\eta/2)\rangle_{1,2}\otimes |\phi(-\eta/2)\rangle_{3,4}\otimes \ldots |\phi(-\eta/2)\rangle_{L-1,L}\,.
\ee
Since the basis vectors $|j_0\rangle$ and $|j_{L+1}\rangle$ are arbitrary, we immediately get
\bea
\hspace{-1.8cm}\Gamma^{t_0}_0(-\eta/2+u) R^{t_0}_{0,1}(-u)R^{t_0}_{0,2}(-u)\ldots R^{t_0}_{0,L}(-u) |\Psi_0\rangle_{1,2,\ldots L}=\nonumber\\
\hspace{-1.4cm}R_{0,1}(-u)R_{0,2}(-u)\ldots R_{0,L}(-u)\Gamma^{t_0}_0(-\eta/2+u) |\Psi_0\rangle_{1,2,\ldots L}\,.
\label{eq:intermediate}
\eea
Let now the parameter $u$ be such that $\Gamma(-u+\eta/2)$ is invertible. Then, we can multiply on the left both sides of \eqref{eq:intermediate} by $\left(\Gamma^{t_0}_0(-\eta/2+u)\right)^{-1}$ and take the trace over the auxiliary space $h_0$. Using the cyclic property of the trace we obtain
\bea
{\rm tr}_0\left\{ R^{t_0}_{0,1}(-u)R^{t_0}_{0,2}(-u)\ldots R^{t_0}_{0,L}(-u) \right\}|\Psi_0\rangle_{1,2,\ldots L}\nonumber\\
={\rm tr}_0\left\{R_{0,1}(-u)R_{0,2}(-u)\ldots R_{0,L}(-u) \right\}|\Psi_0\rangle_{1,2,\ldots L}\,.
\label{eq:auxiliary_eq_1}
\eea
Since the matrix $\Gamma(-u+\eta/2)$ is invertible except for a set of isolated points, Eq.~\eqref{eq:auxiliary_eq_1} holds for all values of $u$ by continuity. Finally, using
\bea
\hspace{-1.8cm}{\rm tr}_0\left\{ R^{t_0}_{0,1}(-u)R^{t_0}_{0,2}(-u)\ldots R^{t_0}_{0,L}(-u)\right\}\nonumber\\
={\rm tr}_0\left\{ \left[R_{0,L}(-u)R_{0,L-1}(-u)\ldots  R_{0,1}(-u)\right]^{t_0} \right\}\nonumber\\
={\rm tr}_0\left\{R_{0,L}(-u)R_{0,L-1}(-u)\ldots  R_{0,1}(-u) \right\}
\,,
\eea
we immediately obtain \eqref{eq:final_result}.

\section{The $Y$-system}\label{sec:y-system}

In this appendix we briefly touch upon another property of the boundary states, which is related to the so-called $Y$-system, an ubiquitous structure of integrability \cite{KuNS11}. The latter is a system of equations for a set of functions in the complex plane. In the case of the XXZ model, the $Y$-system takes the form
\be
\left[ 1+Y_{j+1}(u)\right]\left[ 1+Y_{j-1}(u)\right]=Y_{j}(u+i\eta/2)Y_{j}(u-i\eta/2)\,,
\ee
where for generic values of $\Delta$, one has $j=1,2,\ldots
\infty$. In the framework of quantum quenches, the $Y$-system first
appeared in the study of quenches from the N\'eel state
\cite{WDBF14}. In this case the $Y$-functions are obtained starting
from the Bethe ansatz rapidity distributions of the corresponding
long-time steady state. Subsequently, the same relations were found to
hold more generally for all two-site product states in the XXZ
spin-$1/2$ model \cite{IQNB16,PiVC16,IlQC17,PoVW17}, and for some
initial states in the spin-$1$ chain \cite{PiVC16}. 

An explanation for the $Y$-system in this context was found in
Ref.~\cite{PiPV17}. Using the identification of two-site states with
boundary states reviewed in Sec.~\ref{sec:integrable_boundaries}, the
$Y$-system emerged from the fusion properties of the corresponding
boundary transfer matrices. From the practical point of view, the
existence of a $Y$-system represents a major computational advance,
allowing for a closed-form analytical characterization of the rapidity
distribution functions of the post-quench steady state
\cite{PiVC16,IQNB16,PiPV17}. 

So far, it was not clear why the existence of the $Y$-system should be
expected to imply the presence of the pair structure discussed in
Sec.~\ref{sec:pair_structure}. The results of
Sec.~\ref{sec:integrable_boundaries} gives us further understanding of
this point. Explicitly, we have proven that boundary states, which
were shown to be characterized by a $Y$-system in \cite{PiPV17},
satisfy the condition \eqref{eq:condition_integrability_lattice}. In
turn, the latter implies the pair structure if no fine tuning of the
couplings is made. Hence, in this case the $Y$-system and the pair
structure are seen to have the same origin, rooted once again in
integrability. 

We remark that if the initial state is such that the overlaps
can be factorized algebraically, i.e. they can be written in the form
of
\begin{equation}
  \label{ovfac}
\left|  \langle \{\lambda_j,-\lambda_j\}_{j=1}^{N/2}|\Psi_0
  \rangle\right|^2
=C(\{\lambda_j\}_{N/2})\prod_{j=1}^{N/2} v(\lambda_j)\,,
\end{equation}
such that $v(\lambda)$ is a meromorphic function and $C(\{\lambda_j\}_{N/2})$ is $\mathcal{O}(L^0)$ in the thermodynamic limit, then the resulting
$Y$-functions (as obtained within the Quench Action formalism \cite{CaEs13})
necessarily satisfy the $Y$-system equations. This can be proven using a simple
analytic manipulation of the resulting TBA equations, in analogy with
the methods presented in an early work on $Y$-systems by Al. B. Zamolodchikov
\cite{zam-y}.
The same statement can be proven even for non-integrable quenches,
where the pair structure does not hold: the $Y$-system would still
hold if the overlaps can be factorized as
\begin{equation}
  \label{ovfac2}
\left|  \left\langle \{\lambda_j\}_{j=1}^{N}|\Psi_0\right
  \rangle\right|^2
=C(\{\lambda_j\}_{N/2})\prod_{j=1}^{N} v(\lambda_j)
\end{equation}
It follows that if the $Y$-system does not hold, then
algebraic factorization of the overlaps is not possible.
This argument suggests that the specific form \eqref{ovfac} is a
further characteristic property of integrable initial states.

\section{Generalities on matrix product states} \label{sec:MPS_states}

In this appendix we provide some technical details on MPSs which are useful for our discussion in Sec.~\ref{sec:xxz_case}. In particular, we show that the MPS \eqref{eq:psiK_state} can always be decomposed as in Eq.~\eqref{eq:decomposition_zero_momentum}. 

We start by noting a special property of the state \eqref{eq:psiK_state}. Consider the operator
\be
N=\sum_{j=1}^{2}\tilde{Z}^{(j)}\otimes \bar{\tilde{Z}}^{(j)}\,,
\label{eq:transfer_matrix_condition_MPS}
\ee
where $\bar{\tilde{Z}}^{(j)}$ is the complex conjugate of
$\tilde{Z}^{(j)}$, and where $\tilde{Z}^{(1)}$ and $\tilde{Z}^{(2)}$ are given in \eqref{eq:tilde_zeta_1} and \eqref{eq:tilde_zeta_2}. The eigenvalues of $N$ form pairs with the same
magnitude and different sign. This can be seen by performing a
similarity transformation
using the matrix $C^{(k)}$, which represents a rotation of $\pi/2$
around the $z$-axis. We have
\begin{equation*}
 C^{(k)}S_+^{(k)}(C^{(k)})^{-1}=i S_+^{(k)}\,,\qquad
   C^{(k)}S_-^{(k)}(C^{(k)})^{-1}=-i S_-^{(k)}\,,
\end{equation*}
and
\begin{equation*}
 \left[C^{(k)}\otimes C^{(k)}\right] N\left[(C^{(k)})^{-1}\otimes (C^{(k)})^{-1}\right]=-N\,.
\end{equation*}
It can be checked that the leading eigenvalues have no further degeneracies.
From general theorems regarding MPSs \cite{PVWC06}, this implies that there exist
two projectors 
$\mathcal{P}_{1,2}$ acting in auxiliary space such that $\mathcal{P}_1+\mathcal{P}_2=1$ and
\begin{equation}
\mathcal{P}_1 \tilde{Z}^{(j)}=\tilde{Z}^{(j)} \mathcal{P}_2\,, \qquad  \mathcal{P}_2 \tilde{Z}^{(j)}=\tilde{Z}^{(j)} \mathcal{P}_1, \qquad j=1,2\,.
\label{eq:P_relation}
\end{equation}
In our case, the operators $\mathcal{P}_1$ and $\mathcal{P}_2$ are defined by the matrices in \eqref{projectors}. Following \cite{PVWC06}, one can simply compute
\bea
\ket{\chi_0}&=&\sum_{i_1,\dots,i_L=1}^2{\rm tr}
\left[\tilde{Z}^{(i_1)}  \tilde{Z}^{(i_1)}  \dots \tilde{Z}^{(i_1)}   \right]
\ket{i_1,i_2,\dots,i_L}\nonumber\\
&=&\sum_{i_1,\dots,i_L=1}^2{\rm tr}
\left[ \mathcal{P}_1 \tilde{Z}^{(i_1)}  \tilde{Z}^{(i_2)}  \dots \tilde{Z}^{(i_L)}  \mathcal{P}_1  \right]
\ket{i_1,i_2,\dots,i_L}+\nonumber\\
&+&\sum_{i_1,\dots,i_L=1}^2{\rm tr}
\left[ \mathcal{P}_2\tilde{Z}^{(i_2)}  \tilde{Z}^{(i_2)}  \dots \tilde{Z}^{(i_L)}  \mathcal{P}_2  \right]
\ket{i_1,i_2,\dots,i_L}\,.
\eea
Using now $\mathcal{P}_{1,2}=\mathcal{P}_{1,2}^2$ and \eqref{eq:P_relation}, it is straightforward to recover \eqref{eq:decomposition_zero_momentum}.

\end{document}